\documentstyle[amsfonts,aps]{revtex}

\begin{document}
\author{Roberto Laura}
\address{Departamento de F\'{\i}sica, F.C.E.I.A., Universidad Nacional de Rosario.\\
Instituto de F\'{\i}sica Rosario, CONICET-UNR\\
Av. Pellegrini 250, 2000 Rosario, Argentina.\\
e-mail: laura@ifir.ifir.edu.ar}
\author{Mario Castagnino}
\address{Instituto de Astronom\'{\i}a y F\'{\i}sica del Espacio,\\
Casilla de Correos 67, Sucursal 28,\\
1428 Buenos Aires, Argentina.\\
e-mail: castagni@iafe.uba.ar}
\author{Rodolfo M. Id Betan}
\address{Departamento de F\'{\i}sica, F.C.E.I.A., Universidad Nacional de Rosario.\\
Instituto de F\'{\i}sica Rosario, CONICET-UNR\\
Av. Pellegrini 250, 2000 Rosario, Argentina.\\
e-mail: rodolfo@ifir.ifir.edu.ar}
\title{Perturbative method for generalized spectral decompositions.}
\date{November 1998}
\maketitle

\begin{abstract}
Imposing analytic properties to states and observables we construct a
perturbative method to obtain a generalized biorthogonal system of
eigenvalues and eigenvectors for quantum unstable systems. A decay process
can be described using this generalized spectral decomposition, and the
final generalized state is obtained.
\end{abstract}

\section{Introduction.}

Following previous ideas about the use of analytic continuation in the study
of unstable quantum systems \cite{3}, \cite{1b}, \cite{1c}, \cite{1}, in
papers \cite{2}, \cite{5} and \cite{6} we have developed our own version,
obtaining new results. But, as there were not a perturbative method adapted
to our formalism, we were forced to study only solvable simple models.
Therefore, this paper is devoted to fill this gap introducing a perturbation
method adapted to our formalism. Let us briefly review the problem and the
history of the proposed solution.

The decay of unstable systems in quantum mechanics is usually described by a
Hamiltonian of the form $H=H_{free}+H_{int}$, where $H_{free}$ has a
discrete spectrum imbedded in the continuous spectrum, and $H$ has a
continuous spectrum. In non equilibrium quantum statistical mechanics, the
generator of the time evolution is the Liouville-Von Newmann operator which
in some cases can be written as ${\Bbb L=L}_{free}+{\Bbb L}_{int}$. There is
an overlapping between the continuous and the discrete part of the spectrum
of ${\Bbb L}_{free}$, where in this case the discrete part of the spectrum
contains the zero eigenvalue of the generator of the free evolution. It is
expected that the infinite dimensional degeneration of the zero eigenvalue
to be partially removed by the interaction, so that the zero eigenspace of $%
{\Bbb L}$ turns out to be the equilibrium subspace.

E.g., for the Friedrichs model, a prototype of a decay system in quantum
mechanics, with Hamiltonian 
\[
H=\Omega |1\rangle \langle 1|+\int_0^\infty d\omega \,\omega |\omega \rangle
\langle \omega |+\int_0^\infty d\omega \,V_\omega \left\{ |\omega \rangle
\langle 1|+|1\rangle \langle \omega |\right\} ,\quad \Omega \in {\Bbb R}%
^{+},V_\omega =\overline{V_\omega }, 
\]
it is possible to find conditions on the interaction $V_\omega $ for which
the spectrum of $H$ is ${\Bbb R}^{+}$. The discrete part of the spectrum of $%
H_0=\Omega |1\rangle \langle 1|+\int_0^\infty d\omega \,\omega |\omega
\rangle \langle \omega |$ is eliminated by the interaction. A complete set
of generalized right (left) eigenvectors $|\omega ^{+}\rangle $ ($\langle
\omega ^{+}|$) of the Hamiltonian $H$ with eigenvalue $\omega \in {\Bbb R}%
^{+}$ can be explicitly obtained, thus $H=\int_0^\infty d\omega \,\omega
|\omega ^{+}\rangle \langle \omega ^{+}|$. When $V_\omega \rightarrow 0$, $%
|\omega ^{+}\rangle $ tends to $|\omega \rangle $ and therefore the discrete
eigenvector $|1\rangle $ is not recovered.

E.C.G.Sudarshan et al \cite{3} obtained a generalized spectral decomposition
for the Friedrichs model by considering a Hamiltonian obtained from the
previous expression by replacing the integrals over ${\Bbb R}^{+}$ by
integrals over a curve $\Gamma $ in the lower complex half plane. This
extended Hamiltonian $H_\Gamma $ is meaningful as the generator of the time
evolution for state vectors $\Phi $ for which $\langle \omega |\Phi \rangle $
has a well defined analytic extension to the lower complex half plane. The
spectrum of $H_\Gamma $ is the curve $\Gamma $ and a point $z_0$ located in
the lower half plane between the curve $\Gamma $ and the real axis. When $%
V_\omega \rightarrow 0$, $z_0$ tends to $\Omega $, and the discrete part of
the spectrum of the ''free'' Hamiltonian is recovered (see also reference 
\cite{2}).

The same complex spectral decomposition for the Friedrichs model was also
obtained by T.Petrosky, I.Prigogine and S.Tasaki \cite{3b}, with a
perturbative method using a ''time ordering rule'' by which an small
imaginary part is included to avoid small denominators, with a different
sign according to the type of transition involved.

Latter, I.Antoniou et.al. \cite{4}, developed a formalism to deal with
singular observables and generalized states, which they applied in the
framework of the subdynamics theory. We gave our version of this formalism
in papers \cite{5} and \cite{6}. We also considered singular observables and
generalized states to describe the time evolution of a quantum oscillator
interacting with a field, describing the evolution of the system to
equilibrium in the thermodynamic limit \cite{6}.

The analytic dilation method \cite{ad1}\cite{ad2} is also a well established
procedure to determine the contributions of the singularities to the time
evolution of unstable quantum systems.

In this paper we consider states and observables with suitable analytic
properties (as in papers \cite{2} and \cite{5}), such that the integrals
over the continuous part of the spectrum can be deformed into integrals over
curves in the complex plane. The physical motivation of these analytic
properties is explained in \cite{2}. In this way we can avoid the ill
defined terms involving the differences between discrete and continuous
eigenvalues of the unperturbed generator of the time evolution, and
implement a well defined algorithm to obtain a complete biorthogonal set of
generalized eigenvectors in one to one correspondence with the basis of the
unperturbed problem. This biorthogonal system can be ued to give a
description of the decay process.

The paper is organized as follows:

In section II we present the perturbative algorithm for the case of pure
states, compute the complete set of right and left generalized eigenvectors
and discuss the roll of the complex eigenvalues in the approximated
expressions for the decay process. The method is applied to study the
tunneling through a barrier in a one dimensional problem and to the
Friedrichs model, were the results are compared with the exact solutions of
reference \cite{3}. In section III the case of generalized states and
singular observables for the Friedrichs model is studied with the
perturbative method. In the conclusions of section IV we summarize the
results and discuss the physical interpretation of the complex generalized
spectral decomposition.

\section{Pure States.}

Let us consider a system where the free Hamiltonian $\stackrel{0}{H}$ has a
continuous spectrum plus a single discrete eigenvalue, and there is a
general interaction $\stackrel{1}{H}$: 
\begin{eqnarray}
H &=&\stackrel{0}{H}+\stackrel{1}{H}  \nonumber \\
\stackrel{0}{H} &\doteq &\Omega |1\rangle \langle 1|+\int_0^\infty d\omega
\,\omega |\omega \rangle \langle \omega |\quad \Omega \in {\Bbb R}^{+}, 
\nonumber \\
\stackrel{1}{H} &\doteq &\int_0^\infty d\omega \,V_\omega |\omega \rangle
\langle 1|+\int_0^\infty d\omega \,\overline{V_\omega }|1\rangle \langle
\omega |+\int_0^\infty d\omega \int_0^\infty d\omega ^{\prime }V_{\omega
\omega ^{\prime }}|\omega \rangle \langle \omega ^{\prime }|,\quad V_{\omega
\omega ^{\prime }}=\overline{V_{\omega ^{\prime }\omega }},\quad \Omega \in 
{\Bbb R}^{+},  \label{a7}
\end{eqnarray}
We suppose that the functions $V_\omega $ and $V_{\omega \omega ^{\prime }}$
are endowed with some analytic properties that we will specify below
equation (\ref{a9}). The generalized right (left) eigenvectors $|1\rangle $
and $|\omega \rangle $ ($\langle 1|$ and $\langle \omega |$) of $\stackrel{0%
}{H}$ form a complete biorthogonal set that we can use to expand any vector
state $|\Phi \rangle $, i.e. 
\begin{equation}
\langle 1|1\rangle =1,\qquad \langle \omega |\omega ^{\prime }\rangle
=\delta (\omega -\omega ^{\prime }),\qquad \langle 1|\omega \rangle =\langle
\omega |1\rangle =0,\quad |\Phi \rangle =|1\rangle \langle 1|\Phi \rangle
+\int_0^\infty d\omega \,|\omega \rangle \langle \omega |\Phi \rangle .
\label{a6}
\end{equation}

The amplitude of a state vector $\Phi ,$ observed in the state vector $\Psi $%
, is given by 
\begin{equation}
\langle \Psi |\Phi \rangle =\langle \Psi |1\rangle \langle 1|\Phi \rangle
+\int_0^\infty d\omega \,\langle \Psi |\omega \rangle \langle \omega |\Phi
\rangle ,  \label{a3}
\end{equation}
and therefore 
\[
\langle \Psi |\Phi \rangle =\langle \Psi |I|\Phi \rangle ,\qquad I\doteq
|1\rangle \langle 1|+\int_0^\infty d\omega \,|\omega \rangle \langle \omega
|. 
\]

If we try to construct a complete biorthogonal set of eigenvectors of the
Hamiltonian $H=\stackrel{0}{H}+\stackrel{1}{H}$, depending analytically on
the small interaction parameter which we assume included in $\stackrel{1}{H}$%
, the standard methods of perturbation theory are not applicable, because
the superposition of the discrete and continuous part of the spectrum of $%
\stackrel{0}{H}$ will produce non defined terms in the perturbation
expansion. However, this problem can be avoided if we are satisfied to
obtain information about terms of the form $\langle \Psi |\Phi \rangle $, $%
\langle \Psi |H|\Phi \rangle $ or $\langle \Psi |\exp (-iHt)|\Phi \rangle $,
where $\Phi $ is a vector for which $\varphi (\omega )=\langle \omega |\Phi
\rangle $ has a well defined analytic extension $\varphi (z)$ to the lower
half plane, and $\Psi $ is a vector for which $\psi (\omega )=\langle \omega
|\Psi \rangle $ has a well defined analytic extension $\psi (z)$ to the
upper half plane. In this case the functionals $\langle z|$ and $|z\rangle $
are defined through the analytic expressions\cite{1}\cite{2}: 
\[
\langle z|\Phi \rangle \doteq \varphi (z),\qquad \langle \Psi |z\rangle
\doteq \overline{\psi (\overline{z})}. 
\]
Then, the Cauchy theorem can be used in equation (\ref{a3}) to write 
\begin{equation}
\langle \Psi |\Phi \rangle =\langle \Psi |1\rangle \langle 1|\Phi \rangle
+\int_0^\infty d\omega \,\langle \Psi |\omega \rangle \langle \omega |\Phi
\rangle =\langle \Psi |1\rangle \langle 1|\Phi \rangle +\int_\Gamma
dz\,\langle \Psi |z\rangle \langle z|\Phi \rangle ,  \label{a8}
\end{equation}
where $\Gamma $ is the curve in the lower complex half plane indicated in
Fig. 1. Therefore, for the states $\Phi $ and observables $\Psi $ endowed
with the analyticity properties mentioned above, we can define the new
''identity operator'' $I_\Gamma $ 
\[
I_\Gamma \doteq |1\rangle \langle 1|+\int_\Gamma dz\,|z\rangle \langle
z|,\qquad \langle \Psi |\Phi \rangle =\langle \Psi |I_\Gamma |\Phi \rangle . 
\]
The functionals $|1\rangle $, $\langle 1|$, $|z\rangle $ and $\langle z|$
satisfy 
\begin{equation}
\langle 1|1\rangle =1,\qquad \langle z|z^{\prime }\rangle =\delta _\Gamma
(z-z^{\prime }),\qquad \langle 1|z\rangle =\langle z|1\rangle =0,
\label{a9'}
\end{equation}
where $\delta _\Gamma (z-z^{\prime })$ is the $\delta $ distribution defined
on the curve $\Gamma $ ($\int_\Gamma dz^{\prime }\delta _\Gamma (z-z^{\prime
})\,f(z^{\prime })=f(z)$).

We can also define $H_\Gamma =\stackrel{0}{H}_\Gamma +\stackrel{1}{H}_\Gamma 
$, where 
\begin{equation}
\stackrel{0}{H}_\Gamma \doteq \Omega |1\rangle \langle 1|+\int_\Gamma
dz\,z|z\rangle \langle z|,\quad \stackrel{1}{H}_\Gamma \doteq \int_\Gamma
dz\,V_z|z\rangle \langle 1|+\int_\Gamma dz\,\overline{V_{\overline{z}}}%
|1\rangle \langle z|+\int_\Gamma dz\int_\Gamma dz^{\prime }V_{zz^{\prime
}}|z\rangle \langle z^{\prime }|,  \label{a9}
\end{equation}
where $V_z$, $\overline{V_{\overline{z}}}$ and $V_{zz^{\prime }}$ are the
analytic extensions to the lower half plane of $V_\omega $, $\overline{%
V_\omega }$ and $V_{\omega \omega ^{\prime }}$. Therefore we postulate that
these functions can be analytically extended at least up to the curve $%
\Gamma $. The operators $\stackrel{0}{H}_\Gamma $ and $\stackrel{1}{H}%
_\Gamma $ verify $\langle \Psi |\stackrel{0}{H}|\Phi \rangle =\langle \Psi |%
\stackrel{0}{H}_\Gamma |\Phi \rangle $ and $\langle \Psi |\stackrel{1}{H}%
|\Phi \rangle =\langle \Psi |\stackrel{1}{H}_\Gamma |\Phi \rangle $ for the
restricted class of vectors defined above, and therefore $H_\Gamma $can be
considered as the generator of the time evolution. Being $\Omega $ a real
number, there is no superposition between the discrete and the continuous
part of the spectrum of $\stackrel{0}{H}_\Gamma $.

We can now proceed with a perturbation approach to the right eigenvalue
problem 
\[
H_\Gamma |\Phi \rangle =\lambda |\Phi \rangle . 
\]
By assuming the expansion 
\[
|\Phi \rangle =\,|\stackrel{0}{\Phi }\rangle +|\stackrel{1}{\Phi }\rangle +|%
\stackrel{2}{\Phi }\rangle +\cdot \cdot \cdot \cdot \cdot ,\quad \lambda =\,%
\stackrel{0}{\lambda }+\stackrel{1}{\lambda }+\stackrel{2}{\lambda }+\cdot
\cdot \cdot \cdot \cdot , 
\]
with respect to the small interaction parameter that we suppose contained in 
$\stackrel{1}{H}_\Gamma $, we obtain order by order the set of equations 
\begin{eqnarray}
\left( \stackrel{0}{\lambda }-\stackrel{0}{H}_\Gamma \right) |\stackrel{0}{%
\Phi }\rangle \, &=&0  \label{a10} \\
\left( \stackrel{0}{\lambda }-\stackrel{0}{H}_\Gamma \right) |\stackrel{1}{%
\Phi }\rangle \, &=&\left( \stackrel{1}{H}_\Gamma -\stackrel{1}{\lambda }%
\right) |\stackrel{0}{\Phi }\rangle  \label{a11} \\
\left( \stackrel{0}{\lambda }-\stackrel{0}{H}_\Gamma \right) |\stackrel{2}{%
\Phi }\rangle \, &=&\left( \stackrel{1}{H}_\Gamma -\stackrel{1}{\lambda }%
\right) |\stackrel{1}{\Phi }\rangle -\stackrel{2}{\lambda }\,|\stackrel{0}{%
\Phi }\rangle  \label{a12} \\
&&\cdot \cdot \cdot \cdot \cdot \cdot \cdot \cdot \cdot \cdot \cdot \cdot
\cdot \cdot  \nonumber \\
\left( \stackrel{0}{\lambda }-\stackrel{0}{H}_\Gamma \right) |\stackrel{n}{%
\Phi }\rangle \, &=&\left( \stackrel{1}{H}_\Gamma -\stackrel{1}{\lambda }%
\right) |\stackrel{n-1}{\Phi }\rangle -\stackrel{2}{\lambda }\,|\stackrel{n-2%
}{\Phi }\rangle -\cdot \cdot \cdot -\stackrel{n}{\lambda }\,|\stackrel{0}{%
\Phi }\rangle  \label{a13} \\
&&\cdot \cdot \cdot \cdot \cdot \cdot \cdot \cdot \cdot \cdot \cdot \cdot
\cdot \cdot  \nonumber
\end{eqnarray}

Analogously, for the left eigenvalue problem we have 
\[
\langle \Psi |H_\Gamma =\lambda \langle \Psi |,\quad \langle \Psi
|=\,\langle \stackrel{0}{\Psi }|+\langle \stackrel{1}{\Psi }|+\langle 
\stackrel{2}{\Psi }|+\cdot \cdot \cdot \cdot \cdot ,\quad \lambda =\,%
\stackrel{0}{\lambda }+\stackrel{1}{\lambda }+\stackrel{2}{\lambda }+\cdot
\cdot \cdot \cdot \cdot , 
\]
and we obtain 
\begin{eqnarray}
\langle \stackrel{0}{\Psi }|\left( \stackrel{0}{\lambda }-\stackrel{0}{H}%
_\Gamma \right) &=&0  \label{a14} \\
\langle \stackrel{1}{\Psi }|\left( \stackrel{0}{\lambda }-\stackrel{0}{H}%
_\Gamma \right) &=&\langle \stackrel{0}{\Psi }|\left( \stackrel{1}{H}_\Gamma
-\stackrel{1}{\lambda }\right)  \label{a15} \\
\langle \stackrel{2}{\Psi }|\left( \stackrel{0}{\lambda }-\stackrel{0}{H}%
_\Gamma \right) &=&\langle \stackrel{1}{\Psi }|\left( \stackrel{1}{H}_\Gamma
-\stackrel{1}{\lambda }\right) -\,\langle \stackrel{0}{\Psi }|\,\stackrel{2}{%
\lambda }  \label{a16} \\
&&\cdot \cdot \cdot \cdot \cdot \cdot \cdot \cdot \cdot \cdot \cdot 
\nonumber \\
\langle \stackrel{n}{\Psi }|\left( \stackrel{0}{\lambda }-\stackrel{0}{H}%
_\Gamma \right) &=&\langle \stackrel{n-1}{\Psi }|\left( \stackrel{1}{H}%
_\Gamma -\stackrel{1}{\lambda }\right) -\,\langle \stackrel{n-2}{\Psi }|\,%
\stackrel{2}{\lambda }-\cdot \cdot \cdot -\langle \stackrel{0}{\Psi }|\,%
\stackrel{n}{\lambda }  \label{a17} \\
&&\cdot \cdot \cdot \cdot \cdot \cdot \cdot \cdot \cdot \cdot \cdot 
\nonumber
\end{eqnarray}

It is convenient to define the projectors 
\begin{equation}
P_d\doteq |1\rangle \langle 1|,\qquad P_\Gamma \doteq \int_\Gamma
dz\,|z\rangle \langle z|,\qquad P_z\doteq |z\rangle \langle z|,  \label{a18}
\end{equation}
endowed with the following properties 
\begin{eqnarray}
P_d+P_\Gamma &=&I_\Gamma ,\qquad P_dP_d=P_d,\qquad P_\Gamma P_\Gamma
=P_\Gamma ,\qquad P_zP_{z^{\prime }}=\delta _\Gamma (z-z^{\prime
})\,P_z,\qquad P_dP_\Gamma =P_\Gamma P_d=0  \nonumber \\
P_d\stackrel{0}{H}_\Gamma &=&\stackrel{0}{H}_\Gamma P_d=\Omega P_d,\qquad
P_\Gamma \stackrel{0}{H}_\Gamma =\stackrel{0}{H}_\Gamma P_\Gamma ,\qquad P_z%
\stackrel{0}{H}_\Gamma =\stackrel{0}{H}_\Gamma P_z=z\,P_z.  \label{a19}
\end{eqnarray}
With all these equations in hand we can begin our computations.

\subsection{The discrete spectrum.}

Let us start computing the eigenvalues and eigenvectors of $H_\Gamma $
corresponding to the unperturbed eigenvalue $\Omega $, so we take $|%
\stackrel{0}{\Phi }\rangle =|1\rangle $ and $\stackrel{0}{\lambda }=\Omega $%
. With this choice equation (\ref{a10}) is verified. Then equation (\ref{a11}%
) gives 
\[
\int_\Gamma dz\,|z\rangle (\Omega -z)\langle z|\stackrel{1}{\Phi }\rangle
=\int_\Gamma dz\,|z\rangle \langle z|\stackrel{1}{H}_\Gamma |1\rangle -%
\stackrel{1}{\lambda }|1\rangle , 
\]
and therefore 
\[
\stackrel{1}{\lambda }=0,\qquad P_\Gamma |\stackrel{1}{\Phi }\rangle =\frac 1%
{(\Omega -\stackrel{0}{H}_\Gamma )}P_\Gamma \stackrel{1}{H}_\Gamma |1\rangle
. 
\]
Equation (\ref{a12}), gives 
\[
\int_\Gamma dz\,|z\rangle (\Omega -z)\langle z|\stackrel{2}{\Phi }\rangle =%
\stackrel{1}{H}_\Gamma |\stackrel{1}{\Phi }\rangle -\stackrel{2}{\lambda }%
|1\rangle , 
\]
and therefore 
\[
\stackrel{2}{\lambda }=\langle 1|\stackrel{1}{H}_\Gamma |\stackrel{1}{\Phi }%
\rangle ,\qquad P_\Gamma |\stackrel{2}{\Phi }\rangle =\frac 1{(\Omega -%
\stackrel{0}{H}_\Gamma )}P_\Gamma \stackrel{1}{H}_\Gamma |\stackrel{1}{\Phi }%
\rangle . 
\]

For $|\stackrel{0}{\Phi }\rangle =|1\rangle $ and $\stackrel{0}{\lambda }%
=\Omega $, equations (\ref{a10})-(\ref{a13}) give no information about $%
\langle 1|\stackrel{n}{\Phi }\rangle $ ($n=1,2,...$). Thus, we impose the
condition $\langle 1|\stackrel{n}{\Phi }\rangle =0$ ($n=1,2,...$), namely
the usual choice in quantum perturbation theory. This choice will fix the
normalization of the eigenvector. In summary, up to second order, we obtain
the right eigenvector and eigenvalue 
\begin{eqnarray}
|\Phi _\Omega \rangle &=&|1\rangle +\left( \frac 1{(\Omega -\stackrel{0}{H}%
_\Gamma )}P_\Gamma \stackrel{1}{H}_\Gamma \right) |1\rangle +\left( \frac 1{%
(\Omega -\stackrel{0}{H}_\Gamma )}P_\Gamma \stackrel{1}{H}_\Gamma \right)
^2|1\rangle ,  \nonumber \\
\lambda _\Omega &=&\Omega +\langle 1|\stackrel{1}{H}_\Gamma \frac 1{(\Omega -%
\stackrel{0}{H}_\Gamma )}P_\Gamma \stackrel{1}{H}_\Gamma |1\rangle .
\label{a20'}
\end{eqnarray}

Using the expressions for $\stackrel{0}{H}_\Gamma $ and $\stackrel{1}{H}%
_\Gamma $ given in equations (\ref{a9}), we obtain 
\[
\lambda _\Omega =\Omega +\int_\Gamma dz\,\frac{V_z\overline{V_{\overline{z}}}%
}{\Omega -z}. 
\]
If $V_z$ and $\overline{V_{\overline{z}}}$ are assumed to be analytic
functions of $z$ in the lower half plane, the integral over $\Gamma $ can be
transformed into an integral over ${\Bbb R}^{+}$: 
\begin{eqnarray}
\lambda _\Omega &=&\Omega +\int_0^\infty d\omega \,\frac{V_\omega \overline{%
V_\omega }}{\Omega -\omega +i0}=\Omega +\int_0^\infty d\omega \left\{ -i\pi
\delta (\omega -\Omega )-{\cal P}\left( \frac 1{\omega -\Omega }\right)
\right\} \left| V_\omega \right| ^2=  \nonumber \\
&=&\Omega -\int_0^\infty d\omega {\cal P}\left( \frac 1{\omega -\Omega }%
\right) \left| V_\omega \right| ^2-i\pi \left| V_\Omega \right| ^2,
\label{a20'''}
\end{eqnarray}
where we see that the perturbation of the eigenvalue $\Omega >0$ gives, up
to second order, a complex eigenvalue $\lambda _\Omega $ with negative
imaginary part (see Fig.1). Notice that the imaginary part would not appear
if $\Omega <0$.

Starting from $\langle \stackrel{0}{\Psi }|=\langle 1|$ and $\stackrel{0}{%
\lambda }=\Omega $, equations (\ref{a14})-(\ref{a17}) and the imposed
conditions $\langle \stackrel{n}{\Psi }|1\rangle =0$ ($n=1,2,...$) provide a
well defined algorithm to obtain the left eigenvector. E.g. up to second
order we obtain 
\begin{equation}
\langle \Psi _\Omega |=\langle 1|+\langle 1|\left( \stackrel{1}{H}_\Gamma
P_\Gamma \frac 1{(\Omega -\stackrel{0}{H}_\Gamma )}\right) +\langle 1|\left( 
\stackrel{1}{H}_\Gamma P_\Gamma \frac 1{(\Omega -\stackrel{0}{H}_\Gamma )}%
\right) ^2,  \label{a20''}
\end{equation}
with the same eigenvalue $\lambda _\Omega $. As we are searching for a
complete biorthogonal system, it is necessary at every step to define the
normalized eigenvectors as 
\begin{equation}
|f_\Omega \rangle \doteq |\Phi _\Omega \rangle /\sqrt{\langle \Psi _\Omega
|\Phi _\Omega \rangle },\qquad \langle \widetilde{f}_\Omega |\doteq \langle
\Psi _\Omega |/\sqrt{\langle \Psi _\Omega |\Phi _\Omega \rangle },
\label{a20}
\end{equation}
which satisfy the condition $\langle \widetilde{f}_\Omega |f_\Omega \rangle
=1$.

\subsection{The continuous spectrum.}

Let us now compute the eigenvalues and eigenvectors of $H_\Gamma $
corresponding to the continuous spectrum of $\stackrel{0}{H}_\Gamma $. So we
now consider equations (\ref{a10})-(\ref{a13}) with $|\stackrel{0}{\Phi }%
\rangle =|u\rangle $ and $\stackrel{0}{\lambda }=u$ ($u\in \Gamma $).
Equation (\ref{a10}) is satisfied, and equation (\ref{a11}) gives 
\[
\int_\Gamma dz\,|z\rangle (u-z)\langle z|\stackrel{1}{\Phi }\rangle
+(u-\Omega )|1\rangle \langle 1|\stackrel{1}{\Phi }\rangle =\int_\Gamma
dz\,|z\rangle \langle z|\stackrel{1}{H}_\Gamma |u\rangle +|1\rangle \langle
1|\stackrel{1}{H}_\Gamma |u\rangle -\stackrel{1}{\lambda }\int_\Gamma
dz\,|z\rangle \delta _\Gamma (z-u), 
\]
or equivalently 
\begin{equation}
(u-z)\langle z|\stackrel{1}{\Phi }\rangle =\langle z|\stackrel{1}{H}_\Gamma
|u\rangle -\stackrel{1}{\lambda }\delta _\Gamma (z-u),\qquad (u-\Omega
)\langle 1|\stackrel{1}{\Phi }\rangle =\langle 1|\stackrel{1}{H}_\Gamma
|u\rangle .  \label{a20a}
\end{equation}
If we assume that $V_{zu}$ is an analytic function of the two variables $z$
and $u$, the term $\langle z|\stackrel{1}{H}_\Gamma |u\rangle =V_{zu}$ does
not include a term proportional to $\delta _\Gamma (z-u)$, and therefore
equations (\ref{a20a}) give 
\begin{equation}
\stackrel{1}{\lambda }=0\qquad P_\Gamma |\stackrel{1}{\Phi }\rangle =\frac 1{%
(u+i0-\stackrel{0}{H}_\Gamma )}P_\Gamma \stackrel{1}{H}_\Gamma |u\rangle
,\qquad P_d|\stackrel{1}{\Phi }\rangle =\frac 1{(u-\stackrel{0}{H}_\Gamma )}%
P_d\stackrel{1}{H}_\Gamma |u\rangle  \label{a20c}
\end{equation}

Equation (\ref{a12}) gives 
\[
\int_\Gamma dz\,|z\rangle (u-z)\langle z|\stackrel{2}{\Phi }\rangle
+(u-\Omega )|1\rangle \langle 1|\stackrel{2}{\Phi }\rangle =\int_\Gamma
dz\,|z\rangle \langle z|\stackrel{1}{H}_\Gamma |\stackrel{1}{\Phi }\rangle
+|1\rangle \langle 1|\stackrel{1}{H}_\Gamma |\stackrel{1}{\Phi }\rangle -%
\stackrel{2}{\lambda }\int_\Gamma dz\,|z\rangle \delta _\Gamma (z-u), 
\]
and therefore 
\begin{equation}
(u-z)\langle z|\stackrel{2}{\Phi }\rangle =\langle z|\stackrel{1}{H}_\Gamma |%
\stackrel{1}{\Phi }\rangle -\stackrel{2}{\lambda }\delta _\Gamma
(z-u),\qquad (u-\Omega )\langle 1|\stackrel{2}{\Phi }\rangle =\langle 1|%
\stackrel{1}{H}_\Gamma |\stackrel{1}{\Phi }\rangle .  \label{a20b}
\end{equation}
Using (\ref{a9}) and (\ref{a20c}) we can compute 
\[
\langle z|\stackrel{1}{H}_\Gamma |\stackrel{1}{\Phi }\rangle =\int_\Gamma
dz^{\prime }\frac{V_{zz^{\prime }}V_{z^{\prime }u}}{u+i0-z^{\prime }}+\frac{%
V_z\overline{V_{\overline{z}}}}{u-\Omega }, 
\]
which does not include any factor proportional to $\delta _\Gamma (z-u)$.
Therefore, equations (\ref{a20b}) give 
\[
\stackrel{2}{\lambda }=0\qquad P_\Gamma |\stackrel{2}{\Phi }\rangle =\frac 1{%
(u+i0-\stackrel{0}{H}_\Gamma )}P_\Gamma \stackrel{1}{H}_\Gamma |\stackrel{1}{%
\Phi }\rangle ,\qquad P_d|\stackrel{2}{\Phi }\rangle =\frac 1{(u-\stackrel{0%
}{H}_\Gamma )}P_d\stackrel{1}{H}_\Gamma |\stackrel{1}{\Phi }\rangle 
\]

In summary, up to second order, we obtain the right eigenvector 
\begin{eqnarray}
|f_u\rangle &=&|u\rangle +\left( \frac 1{(u+i0-\stackrel{0}{H}_\Gamma )}%
P_\Gamma \stackrel{1}{H}_\Gamma +\frac 1{(u-\stackrel{0}{H}_\Gamma )}P_d%
\stackrel{1}{H}_\Gamma \right) |u\rangle +  \nonumber \\
&&+\left( \frac 1{(u+i0-\stackrel{0}{H}_\Gamma )}P_\Gamma \stackrel{1}{H}%
_\Gamma +\frac 1{(u-\stackrel{0}{H}_\Gamma )}P_d\stackrel{1}{H}_\Gamma
\right) ^2|u\rangle ,  \label{a21}
\end{eqnarray}
with eigenvalue $\lambda _u=u$. Notice that we have included the factor $i0$%
, namely we make the usual choice of outgoing solutions of scattering
theory, to avoid the continuous-continuous resonances.

Equations (\ref{a14})-(\ref{a17}) with $\langle \stackrel{0}{\Psi }|=\langle
u|$ and $\stackrel{0}{\lambda }=u\in \Gamma $ are used to obtain the
corresponding left eigenvector. Up to second order we obtain the left
eigenvector 
\begin{eqnarray}
\langle \widetilde{f}_u| &=&\langle u|+\langle u|\left( \stackrel{1}{H}%
_\Gamma P_\Gamma \frac 1{(u-i0-\stackrel{0}{H}_\Gamma )}+\stackrel{1}{H}%
_\Gamma P_d\frac 1{(u-\stackrel{0}{H}_\Gamma )}\right) +  \nonumber \\
&&\langle u|\left( \stackrel{1}{H}_\Gamma P_\Gamma \frac 1{(u-i0-\stackrel{0%
}{H}_\Gamma )}+\stackrel{1}{H}_\Gamma P_d\frac 1{(u-\stackrel{0}{H}_\Gamma )}%
\right) ^2.  \label{a22}
\end{eqnarray}

It can be easily verified that the vectors given in equations (\ref{a20})-(%
\ref{a22}) satisfy, {\it up to second order}, the orthogonality relations 
\begin{equation}
\langle \widetilde{f}_\Omega |f_\Omega \rangle =1,\quad \langle \widetilde{f}%
_u|f_{u^{\prime }}\rangle =\delta _\Gamma (u-u^{\prime }),\quad \langle 
\widetilde{f}_\Omega |f_u\rangle =\langle \widetilde{f}_u|f_\Omega \rangle
=0.  \label{a23}
\end{equation}
They also verify, up to second order, the ''completeness relation'' 
\begin{equation}
\langle \Psi |\Phi \rangle =\langle \Psi |I_\Gamma |\Phi \rangle ,\qquad
I_\Gamma \doteq |1\rangle \langle 1|+\int_\Gamma du\,|u\rangle \langle
u|=|f_\Omega \rangle \langle \widetilde{f}_\Omega |+\int_\Gamma
du\,|f_u\rangle \langle \widetilde{f}_u|.  \label{a24}
\end{equation}
For the operators $H$ and $H_\Gamma $ defined in equations (\ref{a7}) and (%
\ref{a9}) we also have 
\begin{equation}
\langle \Psi |H|\Phi \rangle =\langle \Psi |H_\Gamma |\Phi \rangle ,\qquad
H_\Gamma =\lambda _\Omega \,|f_\Omega \rangle \langle \widetilde{f}_\Omega
|+\int_\Gamma du\,\lambda _u\,|f_u\rangle \langle \widetilde{f}_u|,
\label{a24a}
\end{equation}
showing that the eigenvalues and eigenvectors obtained above provide the
spectral decomposition of $H_\Gamma $. It should be stressed that the
expressions for $I_\Gamma $ and $H_\Gamma $ given in equations (\ref{a24})
and (\ref{a24a})are meaningful {\it only} when they are used to compute $%
\langle \Psi |I_\Gamma |\Phi \rangle $ or $\langle \Psi |H_\Gamma |\Phi
\rangle $ for vectors $\Psi $ ($\Phi $) for which $\psi (\omega )=\langle
\omega |\Psi \rangle $ ($\varphi (\omega )=\langle \omega |\Phi \rangle $)
have a well defined analytic extension to the complex upper (lower) half
plane.

\subsection{The time evolution.}

Let us make a first application of the above results. The probability $P_t$
of the state vector $\Phi _t=e^{-iHt}\Phi _0$ at the time $t$ of being
measured in the state $\Psi $ is given by 
\begin{equation}
P_t=\left| \langle \Psi |\exp (-iHt)|\Phi _0\rangle \right| ^2=\left|
\langle \Psi |\exp (-iH_\Gamma t)|\Phi _0\rangle \right| ^2,  \label{a25}
\end{equation}
where 
\begin{equation}
\langle \Psi |\exp (-iH_\Gamma t)|\Phi _0\rangle =\exp (-i\,\lambda _\Omega
\,t)\langle \Psi |f_\Omega \rangle \langle \widetilde{f}_\Omega |\Phi
_0\rangle +\int_\Gamma du\,\exp (-i\,u\,t)\langle \Psi |f_u\rangle \langle 
\widetilde{f}_u|\Phi _0\rangle .  \label{a26}
\end{equation}
If the exact expressions for the generalized eigenvalues and eigenvectors
were known, equations (\ref{a25}) and (\ref{a26}) would give $P_t$ exactly.
An approximated expression will be obtained if eigenvalues and eigenvectors
are given up to some finite order in the perturbation expansion.

Let us consider the survival amplitude of the unstable state $|1\rangle $,
given by 
\[
\langle 1|\exp (-iH_\Gamma t)|1\rangle =\exp (-i\,\lambda _\Omega
\,t)\langle 1|f_\Omega \rangle \langle \widetilde{f}_\Omega |1\rangle
+\int_\Gamma du\,\exp (-i\,u\,t)\langle 1|f_u\rangle \langle \widetilde{f}%
_u|1\rangle . 
\]
For a small coupling between the discrete and the continuous spectrum, i.e.
if in the expression for $\stackrel{1}{H}$ given in equation (\ref{a7})
there is a small multiplicative parameter $\varepsilon $ included in $%
V_\omega $, and using the explicit expressions of the generalized
eigenvectors given in sections IIIA and IIIB, we obtain 
\begin{equation}
\langle 1|f_\Omega \rangle \langle \widetilde{f}_\Omega |1\rangle
=1+O(\varepsilon ^2),\qquad \langle 1|f_u\rangle \langle \widetilde{f}%
_u|1\rangle =O(\varepsilon ^2).  \label{a26b}
\end{equation}
If $\lambda _\Omega $ is computed up to second order from equation (\ref
{a20'''}), we have 
\begin{equation}
\left| \langle 1|\exp (-iHt)|1\rangle \right| ^2\cong \exp (-2\pi V_\Omega
^2\,t).  \label{a27}
\end{equation}
This expression is valid if $\varepsilon ^2\ll 1$ and $\varepsilon ^2t\leq 1$%
, i.e. for small coupling between the discrete and the continuous spectrums
and for not too large times. This is the standard approximated expression
for a decay process which can be found in several text books on quantum
mechanics \cite{7}. The deviations from exponential behavior (Zeno and
Khalfin effects for shot and long times respectively), are not shown in this
approximation. Concerning Zeno effect, notice that the time derivative for $%
t=0$ of expression (\ref{a27}), is of order $\varepsilon ^2$. If we compute
the survival probability keeping the second order terms coming from
equations (\ref{a26b}), we can obtain the Zeno effect, because the initial
time derivative at $t=0$ turns out to be of order $\varepsilon ^4$.

\subsection{Tunneling through a barrier.}

We now consider some particular models. Let us first consider the one
dimensional rectangular well problem with the following Hamiltonian, given
in coordinate representation 
\begin{equation}
\stackrel{0}{H}^{\prime }=-\frac{\hbar ^2}{2\mu }\frac{d^2}{dx^2}+\stackrel{0%
}{V}(x),  \label{t1}
\end{equation}
where 
\begin{equation}
\stackrel{0}{V}(x)= 
\begin{array}{l}
V_0\qquad |x|>a \\ 
0\qquad \;\;|x|<a.
\end{array}
\label{t2}
\end{equation}

The spectral decomposition of $\stackrel{0}{H}^{\prime }$ is well known. The
discrete part of the spectrum has eigenvalues $E_j$ between $0$ and $V_0$.
In coordinate representation, the corresponding normalized eigenvectors $%
\langle x|j\rangle $ are of the form $\exp \left( \mp \sqrt{\frac{2\mu }{%
\hbar ^2}(V_0-E_j)}x\right) $, the minus (plus) sign corresponding to $x>a$ (%
$x<-a$). Therefore the vector states $\langle x|j\rangle $ are concentrated
inside the rectangular well. We are going to consider the simple case in
which the parameters $a$ and $V_0$ verify the condition $a\sqrt{\frac{2\mu
V_0}{\hbar ^2}}<\frac \pi 2$, where there is a single discrete eigenvalue $%
E_1$, and the corresponding eigenvector $\langle x|1\rangle $ is an even
function of the coordinate $x$\cite{Flu}. There is also a continuous
spectrum with generalized eigenvalues $E_k=\frac{\hbar ^2}{2\mu }k^2+V_0$ ($%
-\infty <k<+\infty $). The generalized eigenvectors $|k\rangle $ can be
obtained as the solutions of the Lipmann-Schwinger equation. The vectors $%
|k\rangle $ and $|1\rangle $ form an orthonormal basis 
\begin{equation}
\langle k|k^{\prime }\rangle =\delta (k-k^{\prime }),\qquad \langle
1|1\rangle =1,\qquad \langle 1|k\rangle =\langle k|1\rangle =0,\qquad
I=|1\rangle \langle 1|+\int dk|k\rangle \langle k|.  \label{t3}
\end{equation}

We can now add to $\stackrel{0}{H}^{\prime }$ the potential 
\begin{equation}
\stackrel{1}{V}^{\prime }(x)= 
\begin{array}{l}
0\qquad \quad \;|x|<b \\ 
-V_1\qquad |x|>b,
\end{array}
\label{t4}
\end{equation}
with $b\gg a$ and $0<V_1<V_0$. Thus, a barrier of height $V_0-V_1$ and
length $b-a$ appears. The potential $\stackrel{1}{V}^{\prime }$ can be
written in terms of the basis given in (\ref{t3}) 
\[
\stackrel{1}{V}^{\prime }=|1\rangle V_{11}\langle 1|+\int dk^{\prime
}|1\rangle V_{1k^{\prime }}\langle k^{\prime }|+\int dkV_{k1}|k\rangle
\langle 1|+\int dk\int dk^{\prime }|k\rangle \widetilde{V}_{kk^{\prime
}}\langle k^{\prime }|, 
\]
where 
\[
V_{11}=-V_1\left[ \int_{-\infty }^{-b}dx\langle 1|x\rangle \langle
x|1\rangle +\int_{+b}^{+\infty }dx\langle 1|x\rangle \langle x|1\rangle
\right] , 
\]
\[
V_{1k}=\overline{V_{k1}}=-V_1\left[ \int_{-\infty }^{-b}dx\langle 1|x\rangle
\langle x|k\rangle +\int_{+b}^{+\infty }dx\langle 1|x\rangle \langle
x|k\rangle \right] , 
\]
\[
\widetilde{V}_{kk^{\prime }}=-V_1\delta (k-k^{\prime })+V_{kk^{\prime
}},\qquad V_{kk^{\prime }}=V_1\int_{-b}^{+b}dx\langle k|x\rangle \langle
x|k^{\prime }\rangle . 
\]
Due to the exponential decrease of $\langle x|1\rangle $ when $x\rightarrow
\pm \infty $, together with the assumption $b\gg a$, it can be expected that 
$V_{11}$ and $V_{1k}$ be small.

The total Hamiltonian $H=\stackrel{0}{H}^{\prime }+\stackrel{1}{V}^{\prime }$
can be also given by $H=\stackrel{0}{H}+\stackrel{1}{H}$ where 
\begin{eqnarray*}
\stackrel{0}{H} &=&(E_1+V_{11})|1\rangle \langle 1|+\int_{-\infty }^{+\infty
}dk(E_k-V_1)|k\rangle \langle k|, \\
\stackrel{1}{H} &=&\int_{-\infty }^{+\infty }dk^{\prime }|1\rangle
V_{1k^{\prime }}\langle k^{\prime }|+\int_{-\infty }^{+\infty
}dkV_{k1}|k\rangle \langle 1|+\int_{-\infty }^{+\infty }dk\int_{-\infty
}^{+\infty }dk^{\prime }|k\rangle V_{kk^{\prime }}\langle k^{\prime }|.
\end{eqnarray*}

Our interest is to describe the time evolution of the unstable state $%
|1\rangle $, which is an even function in coordinate representation. Only
the even part of the Hamiltonian is relevant to this process 
\begin{eqnarray*}
\stackrel{0}{H}_{even} &=&(E_1+V_{11})|1\rangle \langle 1|+\int_0^{+\infty
}dk(E_k-V_1)|k_{even}\rangle \langle k_{even}|, \\
\stackrel{1}{H}_{even} &=&\int_0^{+\infty }dk^{\prime }|1\rangle
V_{1k^{\prime }}\langle k_{even}^{\prime }|+\int_0^{+\infty
}dkV_{k1}|k_{even}\rangle \langle 1|+\int_0^{+\infty }dk\int_0^{+\infty
}dk^{\prime }|k_{even}\rangle V_{kk^{\prime }}\langle k_{even}^{\prime }|,
\end{eqnarray*}
where $|k_{even}\rangle =\frac 1{\sqrt{2}}(|k\rangle +|-k\rangle )$. If the
integrals over ${\Bbb R}^{+}$ are deformed into integrals over a curve $%
\Gamma $ in the lower complex half plane, and if $V_{11}$ and $V_{1k}$ are
small, we can use the perturbation method of the beginning of this section.
The correction to the discrete eigenvalue $(E_1+V_{11})$ of $\stackrel{0}{H}%
_{even}$ due to the interaction is 
\begin{eqnarray*}
&&\langle 1|\stackrel{1}{H}_{even}\frac 1{E_1+V_{11}-\stackrel{0}{H}%
_{even}+i0}\stackrel{1}{H}_{even}|1\rangle \\
&=&\int_0^{+\infty }\frac{V_{1k}V_{k1}}{(E_1+V_{11})-(\frac{\hbar ^2k^2}{%
2\mu }+V_0-V_1)+i0}dk= \\
&=&-i\pi \int_0^{+\infty }V_{1k}V_{k1}\delta \left[ (E_1+V_{11})-(\frac{%
\hbar ^2k^2}{2\mu }+V_0-V_1)\right] dk-\int_0^{+\infty }V_{1k}V_{k1}{\cal P}%
\left[ \frac 1{(E_1+V_{11})-(\frac{\hbar ^2k^2}{2\mu }+V_0-V_1)}\right] dk.
\end{eqnarray*}

If $(E_1+V_{11})>(V_0-V_1)$ the corrected eigenvalue has an imaginary part $%
-i\frac{\pi \mu }{\hbar ^2\widetilde{k}}V_{1\widetilde{k}}V_{\widetilde{k}1}$%
, where 
\[
\widetilde{k}=\sqrt{\frac{2\mu }{\hbar ^2}\left[
(E_1+V_{11})-(V_0-V_1)\right] }, 
\]
and the survival probability of the unstable state $|1\rangle $ is
approximately given by 
\[
\left| \langle 1|\exp (-iHt)|1\rangle \right| ^2\cong \exp \left( -\frac{%
2\pi \mu }{\hbar ^2\widetilde{k}}V_{1\widetilde{k}}V_{\widetilde{k}1}\right) 
\]

\subsection{Friedrichs model.}

We will finally check our method with the well known solvable Friedrichs
model. This model can be obtained as a special case of equation (\ref{a7}),
with $V_{\omega \omega ^{\prime }}=0$. The Hamiltonian is $H=\stackrel{0}{H}+%
\stackrel{1}{H}$, where 
\begin{equation}
\stackrel{0}{H}=\Omega |1\rangle \langle 1|+\int_0^\infty d\omega \,\omega
|\omega \rangle \langle \omega |,\qquad \stackrel{1}{H}=\int_0^\infty
d\omega \,\left\{ V_\omega |\omega \rangle \langle 1|+\overline{V_\omega }%
\,|1\rangle \langle \omega |\right\} ,\qquad \Omega \in {\Bbb R}^{+}
\label{b1}
\end{equation}

As we considered in the previous section, the variable $\omega \in {\Bbb R}%
^{+}$ can be transformed into a variable $z\in \Gamma $, being $\Gamma $ the
curve in the lower complex half plane shown in Fig.1. The corresponding
Hamiltonian $H_\Gamma $ is given by 
\begin{equation}
H_\Gamma =\stackrel{0}{H}_\Gamma +\stackrel{1}{H}_\Gamma ,\qquad \stackrel{0%
}{H}_\Gamma =\Omega |1\rangle \langle 1|+\int_\Gamma dz\,z|z\rangle \langle
z|,\qquad \stackrel{1}{H}_\Gamma =\int_\Gamma dz\,V_z|z\rangle \langle
1|+\int_\Gamma dz\,\overline{V_{\overline{z}}}|1\rangle \langle z|,
\label{b6}
\end{equation}
where $V_z$ and $\overline{V_{\overline{z}}}$ are the analytic extensions of 
$V_\omega $ and $\overline{V_\omega }$.

It is possible to obtain exact solutions for the right and left eigenvalue
problems $H_\Gamma |\Phi \rangle =\lambda |\Phi \rangle $ and $\langle \Psi
|H_\Gamma =\lambda \langle \Psi |$, where $|\Phi \rangle =|1\rangle \langle
1|\Phi \rangle +\int_\Gamma dz\,z|z\rangle \langle z|\Phi \rangle $ and $%
\langle \Psi |=\langle \Psi |1\rangle \langle 1|+\int_\Gamma dz\langle \Psi
|z\rangle \langle z|$. If for the sake of simplicity 
\begin{equation}
\eta (\lambda )\doteq \lambda -\Omega -\int_\Gamma dz\,\frac{V_z\overline{V_{%
\overline{z}}}}{\lambda -z}  \label{ap1}
\end{equation}
verifies $\eta (\lambda _\Omega )=0$ for just a single complex number $%
\lambda _\Omega $ located in the region of the complex plane between ${\Bbb R%
}^{+}$ and the curve $\Gamma $, the following spectral decomposition is
obtained \cite{3} 
\[
H_\Gamma =\lambda _\Omega \,|f_\Omega \rangle \langle \widetilde{f}_\Omega
|+\int_\Gamma du\,u\,|f_u\rangle \langle \widetilde{f}_u|, 
\]
where 
\begin{eqnarray}
|f_\Omega \rangle &=&\frac 1{\sqrt{\eta ^{\prime }(\lambda _\Omega )}}%
\left\{ |1\rangle +\int_\Gamma dz\,\frac{V_z}{\lambda _\Omega -z}\,|z\rangle
\right\}  \nonumber \\
\langle \widetilde{f}_\Omega | &=&\frac 1{\sqrt{\eta ^{\prime }(\lambda
_\Omega )}}\left\{ \langle 1|+\int_\Gamma dz\,\frac{\overline{V_{\overline{z}%
}}}{\lambda _\Omega -z}\,\langle z|\right\}  \nonumber \\
|f_u\rangle &=&|u\rangle +\frac{\overline{V_{\overline{u}}}}{\eta
(u+i\varepsilon )}\left\{ |1\rangle +\int_\Gamma dz\,\frac{V_z}{(u+i0)-z}%
\,|z\rangle \,\right\}  \nonumber \\
\langle \widetilde{f}_u| &=&\langle u|+\frac{V_u}{\eta (u-i0)}\left\{
\langle 1|+\int_\Gamma dz\,\frac{\overline{V_{\overline{z}}}}{(u-i0)-z}%
\,\langle z|\right\} ,\quad u\in \Gamma .  \label{ap2}
\end{eqnarray}
These generalized eigenvectors form a complete biorthogonal system \cite{3},
in the sense given in equations (\ref{a23}) and (\ref{a24}).

Using equations (\ref{a20'})-(\ref{a20}) we can obtain the approximated
eigenvectors

\begin{eqnarray}
|f_\Omega \rangle &\cong &|1\rangle +\int_\Gamma dz\,\frac{V_z}{\Omega -z}%
\,|z\rangle -\frac 12\int_\Gamma dz\,\frac{V_z\overline{V_{\overline{z}}}}{%
(\Omega -z)^2}\,|1\rangle  \nonumber \\
\langle \widetilde{f}_\Omega | &\cong &\langle 1|+\int_\Gamma dz\,\frac{%
\overline{V_{\overline{z}}}}{\Omega -z}\,\langle z|-\frac 12\int_\Gamma dz\,%
\frac{\overline{V_{\overline{z}}}V_z}{(\Omega -z)^2}\,\langle 1|,  \label{b2}
\end{eqnarray}
with the eigenvalue 
\begin{equation}
\lambda _\Omega \cong \Omega +\int_\Gamma dz\,\frac{V_z\overline{V_{%
\overline{z}}}}{\Omega -z}=\Omega -\int_0^\infty d\omega {\cal P}\frac 1{%
\omega -\Omega }\left| V_\omega \right| ^2-i\pi \left| V_\Omega \right| ^2.
\label{b2'}
\end{equation}

Using equations (\ref{a21}) and (\ref{a22}) we also obtain, up to second
order

\begin{eqnarray}
|f_u\rangle &\cong &|u\rangle +\frac{\overline{V_{\overline{u}}}}{u-\Omega }%
|1\rangle +\frac{\overline{V_{\overline{u}}}}{u-\Omega }\int_\Gamma dz\,%
\frac{V_z}{(u-z+i\varepsilon )}|z\rangle  \nonumber \\
\langle \widetilde{f}_u| &\cong &\langle u|+\frac{V_u}{u-\Omega }\langle 1|+%
\frac{V_u}{u-\Omega }\int_\Gamma dz\,\frac{\overline{V_{\overline{z}}}}{%
(u-z-i\varepsilon )}\langle z|,  \label{b3}
\end{eqnarray}
with eigenvalue $\lambda _u=u$ ($u\in \Gamma $).

In order to compare these results with the approximate eigenvalues and
eigenvectors obtained in equations (\ref{b2})-(\ref{b3}), we need an
expansion of the exact expressions (\ref{ap2}) in powers of the interaction
parameter. We can use equation (\ref{ap1}) to obtain the proper eigenvalue $%
\lambda _\Omega $, satisfying $\eta (\lambda _\Omega )=0$, from the
iterative equation 
\[
\stackrel{n+1}{\lambda _\Omega }=\Omega +\int_\Gamma dz\,\frac{V_z\overline{%
V_{\overline{z}}}}{\left( \stackrel{n}{\lambda _\Omega }-z\right) },\qquad
,n=0,1,2,...,\qquad \stackrel{0}{\lambda }=\Omega . 
\]
For small values of the interaction parameter, the secuence $\stackrel{n}{%
\lambda _\Omega }$ converge to a single solution $\lambda _\Omega $ of $\eta
(\lambda _\Omega )=0$. Up to second order we have 
\begin{equation}
\lambda _\Omega \cong \Omega +\int_\Gamma dz\,\frac{V_z\overline{V_{%
\overline{z}}}}{\Omega -z}.  \label{ap3}
\end{equation}
Up to second order in the interaction parameter, we also have 
\begin{equation}
\frac 1{\sqrt{\eta ^{\prime }(\lambda _\Omega )}}\cong 1-\frac 12\int_\Gamma
dz\,\frac{V_z\overline{V_{\overline{z}}}}{(\Omega -z)^2},\quad \frac 1{%
\lambda _\Omega -z}=\frac 1{\Omega -z}+O(V^2),\quad \frac 1{\eta (u\pm
i\varepsilon )}=\frac 1{u-\Omega }+O(V^2)  \label{ap4}
\end{equation}
Replacing (\ref{ap3}) and (\ref{ap4}) in (\ref{ap2}) we obtain the
approximated expressions (\ref{b2})-(\ref{b3}), so our approximated
algorithm reproduces the results of the solvable model.

\section{Generalized states and singular observables.}

Up to now we have only considered pure states and observables which can be
represented by projectors constructed with normalized vectors. The extension
of the formalism to mixed states, represented by density operators of the
form $\widehat{\rho }=\sum_jp_j|\Phi _j\rangle \langle \Phi _j|$ ($\langle
\Phi _j|\Phi _j\rangle =1$, $p_j\geq 0$, $\sum_jp_j=1$) and observables of
the same form ($\widehat{O}=\sum_jO_j|\Psi _j\rangle \langle \Psi _j|$) is
straightforward. However, singular observables as $\int_{\omega _1}^{\omega
_2}d\omega \,|\omega \rangle \langle \omega |$, which are not operators of
the form just mentioned, normally appear, and they can not be considered in
the framework of the perturbation method presented in the previous section
or its straightforward generalization. In this case, it is possible to
consider the states as functionals acting on the set of observables
represented by operators including singular terms \cite{4}\cite{5}\cite{6}.
These generalized states are also useful to study the approach to
equilibrium in the thermodynamic limit \cite{6}. The corresponding
perturbation method is therefore a non trivial generalization of the one in
section II. This generalization will be the subject of this section and, as
we will see, contains new and interesting features.

Let us consider the Friedrichs model, presented in section II$.$ The
Hamiltonian is given by equation (\ref{b1}) and we assume, for simplicity,
that $V_\omega =\overline{V_\omega }$.

For this model we are going to consider the class of observables which can
be represented by operators of the form 
\begin{equation}
O=O_1|1\rangle \langle 1|+\int_0^\infty d\omega \,O_\omega \,|\omega \rangle
\langle \omega |+\int_0^\infty d\omega \int_0^\infty d\omega ^{\prime
}\,O_{\omega \omega ^{\prime }}\,|\omega \rangle \langle \omega ^{\prime
}|+\int_0^\infty d\omega \,O_{\omega 1}\,|\omega \rangle \langle
1|+\int_0^\infty d\omega ^{\prime }\,O_{1\omega ^{\prime }}\,|1\rangle
\langle \omega ^{\prime }|  \label{la2}
\end{equation}
where $O_1=\overline{O_1}$, $O_\omega =\overline{O_\omega }$, $O_{\omega
\omega ^{\prime }}=\overline{O_{\omega ^{\prime }\omega }}$, and $O_{\omega
1}=\overline{O_{1\omega }}$, being $O_\omega $, $O_{\omega \omega ^{\prime
}} $, $O_{\omega 1}$ and $O_{1\omega ^{\prime }}$ regular functions of the
variables $\omega $ and $\omega ^{\prime }$. The second term of the r.h.s.
of the last equation is a singular operator.

In the Heisemberg representation, the time evolution of an observable is
given by \footnote{%
Superoperators, i.e. operators acting on operators, will be written in
blackboard bold types.} 
\[
-i\frac d{dt}O_t={\Bbb L}O_t,\qquad {\Bbb L}O\doteq [H,O], 
\]
where ${\Bbb L}$ is the Liouville-Von Newmann superoperator.

It is convenient to introduce the notation $|O)\doteq O$ and to define the
generalized operators 
\begin{equation}
|1)\doteq |1\rangle \langle 1|,\qquad |\omega )\doteq |\omega \rangle
\langle \omega |,\qquad |\omega \omega ^{\prime })\doteq |\omega \rangle
\langle \omega ^{\prime }|,\qquad |\omega 1)\doteq |\omega \rangle \langle
1|,\qquad |1\omega ^{\prime })\doteq |1\rangle \langle \omega ^{\prime }|.
\label{la3}
\end{equation}
Therefore we can write 
\begin{equation}
|O)=O_1|1)+\int_0^\infty d\omega \,O_\omega \,|\omega )+\int_0^\infty
d\omega \int_0^\infty d\omega ^{\prime }\,O_{\omega \omega ^{\prime
}}\,|\omega \omega ^{\prime })+\int_0^\infty d\omega \,O_{\omega 1}\,|\omega
1)+\int_0^\infty d\omega ^{\prime }\,O_{1\omega ^{\prime }}\,|1\omega
^{\prime }).  \label{la4}
\end{equation}

When the states of the system are represented by the usual density operators 
$\widehat{\rho }=\widehat{\rho }^{\dagger }$, the mean value of an
observable $O$ is given by $\langle O\rangle _{\widehat{\rho }}=Tr(\widehat{%
\rho }^{\dagger }O)$, and any density operator has the following well known
properties 
\begin{equation}
Tr(\widehat{\rho }^{\dagger })=Tr(\widehat{\rho }^{\dagger }I)=1,\quad Tr(%
\widehat{\rho }^{\dagger }[\alpha A+\beta B]=\alpha Tr(\widehat{\rho }%
^{\dagger }A)+\beta Tr(\widehat{\rho }^{\dagger }B),\quad Tr(\widehat{\rho }%
^{\dagger }B^{\dagger })=\overline{Tr(\widehat{\rho }^{\dagger }B)},
\label{la5}
\end{equation}
where $I$ is the identity operator ($I\doteq |1\rangle \langle
1|+\int_0^\infty d\omega |\omega \rangle \langle \omega |=|1)+\int_0^\infty
d\omega |\omega )$).

As we are going to consider the possibility to have more general states than
the usual density operators $\widehat{\rho }$, we will consider generalized
states as functionals $\rho $ acting on the space of observables defined in
equation (\ref{la4}). The action of a state functional $\rho $ on an
observable $O$ will be denoted by $(\rho |O)$ and it gives the mean value $%
\langle O\rangle _\rho $ of the observable in the state $\rho $ ($\langle
O\rangle _\rho \doteq (\rho |O)$). For any state functional $\rho $ we
assume the following properties 
\begin{equation}
(\rho |I)=1,\qquad (\rho |\alpha A+\beta B)=\alpha (\rho |A)+\beta (\rho
|B),\qquad (\rho |B^{\dagger })=\overline{(\rho |B)},  \label{la6}
\end{equation}
which are the generalization of the properties for the density operators $%
\widehat{\rho }$ given in equations (\ref{la5}).

For the class of observables given in equation (\ref{la4}) it is convenient
to {\it define} the functionals $(1|$, $(\omega |$, $(\omega \omega ^{\prime
}|$, $(\omega 1|$ and $(1\omega ^{\prime }|$, in such a way that 
\begin{equation}
(1|O)=O_1,\quad (\omega |O)=O_\omega ,\quad (\omega \omega ^{\prime
}|O)=O_{\omega \omega ^{\prime }},\quad (\omega 1|O)=O_{\omega 1},\quad
(1\omega ^{\prime }|O)=O_{1\omega ^{\prime }}.  \label{la7}
\end{equation}

Using (\ref{la4}), (\ref{la6}) and (\ref{la7}) we obtain 
\begin{eqnarray}
(\rho |O) &=&(\rho |1)(1|O)+\int_0^\infty d\omega (\rho |\omega )(\omega
|O)+\int_0^\infty d\omega \int_0^\infty d\omega ^{\prime }(\rho |\omega
\omega ^{\prime })(\omega \omega ^{\prime }|O)+  \nonumber \\
&&+\int_0^\infty d\omega (\rho |\omega 1)(\omega 1|O)+\int_0^\infty d\omega
^{\prime }(\rho |1\omega ^{\prime })(1\omega ^{\prime }|O).  \label{la8}
\end{eqnarray}
From this equation we obtain 
\[
(\rho |O)=(\rho |\,{\Bbb I\,}|O), 
\]
where ${\Bbb I}$ is the identity superoperator 
\begin{equation}
{\Bbb I}\doteq |1)(1|+\int_0^\infty d\omega |\omega )(\omega |+\int_0^\infty
d\omega \int_0^\infty d\omega ^{\prime }|\omega \omega ^{\prime })(\omega
\omega ^{\prime }|+\int_0^\infty d\omega |\omega 1)(\omega 1|+\int_0^\infty
d\omega ^{\prime }|1\omega ^{\prime })(1\omega ^{\prime }|.  \label{la9}
\end{equation}
The generalized state functionals and observables defined in (\ref{la3}) and
(\ref{la7}) satisfy the orthogonality conditions 
\[
(1|1)=1,\quad (\omega |\omega ^{\prime })=\delta (\omega -\omega ^{\prime
}), 
\]
\[
(\omega \omega ^{\prime }|\xi \xi ^{\prime })=\delta (\omega -\xi )\delta
(\omega ^{\prime }-\xi ^{\prime }),\quad (\omega 1|\xi 1)=\delta (\omega
-\xi ),\quad (1\omega ^{\prime }|1\xi ^{\prime })=\delta (\omega ^{\prime
}-\xi ^{\prime }), 
\]
\[
(1|\omega )=(1|\xi \xi ^{\prime })=(1|\xi 1)=(1|1\xi ^{\prime })=0,\quad
(\omega \omega ^{\prime }|1)=(\omega \omega ^{\prime }|\xi )=(\omega \omega
^{\prime }|\xi 1)=(\omega \omega ^{\prime }|1\xi ^{\prime })=0, 
\]
\begin{equation}
(\omega 1|1)=(\omega 1|\xi )=(\omega 1|\xi \xi ^{\prime })=(\omega 1|1\xi
^{\prime })=0,\quad (1\omega ^{\prime }|1)=(1\omega ^{\prime }|\xi
)=(1\omega ^{\prime }|\xi \xi ^{\prime })=(1\omega ^{\prime }|\xi 1)=0.
\label{la9'}
\end{equation}

Defining the superoperators $\stackrel{0}{\Bbb L}$ and $\stackrel{1}{\Bbb L}$
by 
\[
\stackrel{0}{\Bbb L}O\doteq [\stackrel{0}{H},O],\qquad \stackrel{1}{\Bbb L}%
O\doteq [\stackrel{1}{H},O],\qquad {\Bbb L}=\stackrel{0}{\Bbb L}+\stackrel{1}%
{\Bbb L}, 
\]
and using the generalized operators and functionals defined in equations (%
\ref{la3}) and (\ref{la7}), we have 
\begin{eqnarray}
\stackrel{0}{\Bbb L} &=&\int d\omega ^{\prime }(\Omega -\omega ^{\prime
})|1\omega ^{\prime })(1\omega ^{\prime }|+\int d\omega (\omega -\Omega
)|\omega 1)(\omega 1|+\int d\omega \int d\omega ^{\prime }(\omega -\omega
^{\prime })|\omega \omega ^{\prime })(\omega \omega ^{\prime }|,
\label{la10} \\
\stackrel{1}{\Bbb L} &=&\int d\omega V_\omega \left[ |\omega 1)-|1\omega
)\right] (1|+\int d\omega V_\omega \left[ |1\omega )-|\omega 1)\right]
(\omega |+\int d\omega \left[ -V_\omega |1)+\int d\omega ^{\prime }V_{\omega
^{\prime }}|\omega ^{\prime }\omega )\right] (1\omega |+  \nonumber \\
&&+\int d\omega \left[ V_\omega |1)-\int d\omega ^{\prime }V_{\omega
^{\prime }}|\omega \omega ^{\prime })\right] (\omega 1|+\int d\omega \int
d\omega ^{\prime }\left[ V_\omega |1\omega ^{\prime })-V_{\omega ^{\prime
}}|\omega 1)\right] (\omega \omega ^{\prime }|.  \label{la11}
\end{eqnarray}

From equation (\ref{la10}) and the orthogonality conditions (\ref{la9'}),
the generalized states and observables defined in equations (\ref{la7}) and (%
\ref{la3}) form a complete biorthogonal set of generalized eigenvectors of
the ''free'' superoperator $\stackrel{0}{\Bbb L}$%
\begin{eqnarray}
\stackrel{0}{\Bbb L}|1) &=&\stackrel{0}{\Bbb L}|\omega )=0,\quad \stackrel{0}%
{\Bbb L}|1\omega ^{\prime })=(\Omega -\omega ^{\prime })|1\omega ^{\prime
}),\quad \stackrel{0}{\Bbb L}|\omega 1)=(\omega -\Omega )|\omega 1),\quad 
\stackrel{0}{\Bbb L}|\omega \omega ^{\prime })=(\omega -\omega ^{\prime
})|\omega \omega ^{\prime }),  \nonumber \\
(1|\stackrel{0}{\Bbb L} &=&(\omega |\stackrel{0}{\Bbb L}=0,\quad (1\omega
^{\prime }|\stackrel{0}{\Bbb L}=(\Omega -\omega ^{\prime })(1\omega ^{\prime
}|,\quad (\omega 1|\stackrel{0}{\Bbb L}=(\omega -\Omega )(\omega 1|,\quad
(\omega \omega ^{\prime }|\stackrel{0}{\Bbb L}=(\omega -\omega ^{\prime
})(\omega \omega ^{\prime }|,  \label{la11'}
\end{eqnarray}

If we try to construct a complete biorthogonal set of eigenvectors of the
Liouville-Von Newmann superoperator ${\Bbb L}=\stackrel{0}{\Bbb L}+\stackrel{%
1}{\Bbb L}$, depending analytically on the small interaction parameter which
we assume included in $\stackrel{1}{\Bbb L}$, the standard methods of
perturbation theory are not applicable: the superposition of the discrete
and continuous parts of the spectrum of $\stackrel{0}{\Bbb L}$ will produce
non defined terms in the perturbation expansion.

However, this problem can be avoided if we are satisfied, as in the pure
state case, to obtain information about terms of the form $(\rho |O)$ or $%
(\rho |{\Bbb L}|O)$, considering states $\rho $ and observables $O$ for
which $(\rho |\omega \omega ^{\prime })$, $(\rho |\omega 1)$, $(\rho
|1\omega ^{\prime })$, $(\omega \omega ^{\prime }|O)$, $(\omega 1|O)$ and $%
(1\omega ^{\prime }|O)$ have well defined analytic extensions $(\rho
|zz^{\prime })$, $(\rho |z1)$, $(\rho |1z^{\prime })$, $(zz^{\prime }|O)$, $%
(z1|O)$ and $(1z^{\prime }|O)$ to complex values of $z$ ($z^{\prime }$) on
the upper (lower) complex half plane. We also consider states $\rho $ and
observables $O$ for which $(\rho |\omega )$ and $(\omega |O)$ have well
defined analytic extensions $(\rho |z)$ and $(z|O)$ for complex values of $z$
near ${\Bbb R}^{+}$ in the upper {\it and} in the lower complex half planes.
The physical bases of these analyticity properties are given in papers \cite
{5} and \cite{6}.

For these states and observables, the Cauchy theorem can be used in equation
(\ref{la8}) to write 
\begin{eqnarray}
(\rho |O) &=&(\rho |1)(1|O)+\int_0^\infty d\omega (\rho |\omega )(\omega
|O)+\int_\Gamma dz\int_{\overline{\Gamma }}dz^{\prime }(\rho |zz^{\prime
})(zz^{\prime }|O)+  \nonumber \\
&&+\int_\Gamma dz(\rho |z1)(z1|O)+\int_{\overline{\Gamma }}dz^{\prime }(\rho
|1z^{\prime })(1z^{\prime }|O),  \label{la13}
\end{eqnarray}
where $\Gamma $ is a curve in the lower complex half plane, as shown in Fig.
1, and $\overline{\Gamma }$ is the conjugate curve in the upper complex half
plane. Therefore 
\begin{equation}
(\rho |O)=(\rho |{\Bbb I}_{ext}|O),\quad {\Bbb I}_{ext}\doteq |1)(1|+\int
d\omega |\omega )(\omega |+\int_{\overline{\Gamma }}dz\int_\Gamma dz^{\prime
}|zz^{\prime })(zz^{\prime }|+\int_{\overline{\Gamma }}dz|z1)(z1|+\int_%
\Gamma dz^{\prime }|1z^{\prime })(1z^{\prime }|.  \label{la15}
\end{equation}
If in the expression for ${\Bbb I}_{ext}$ given in equation (\ref{la15}),
the term $\int d\omega |\omega )(\omega |$ is replaced by $\int_{\overline{%
\Gamma }}dz|z)(z|$ or by $\int_\Gamma dz^{\prime }|z^{\prime })(z^{\prime }|$%
, the identity $(\rho |O)=(\rho |{\Bbb I}_{ext}|O)$ is also verified for the
restricted classes of states and observables with the well defined analytic
extensions defined above.

We can also define ${\Bbb L}_{ext}=\stackrel{0}{\Bbb L}_{ext}+\stackrel{1}%
{\Bbb L}_{ext}$, where 
\begin{eqnarray}
\stackrel{0}{\Bbb L}_{ext} &\doteq &\int_\Gamma dz^{\prime }(\Omega
-z^{\prime })|1z^{\prime })(1z^{\prime }|+\int_{\overline{\Gamma }%
}dz(z-\Omega )|z1)(z1|+\int_{\overline{\Gamma }}dz\int_\Gamma dz^{\prime
}(z-z^{\prime })|zz^{\prime })(zz^{\prime }|,  \label{la16} \\
\stackrel{1}{\Bbb L}_{ext} &\doteq &\int_{\overline{\Gamma }%
}dzV_z|z1)(1|-\int_\Gamma dz^{\prime }V_{z^{\prime }}|1z^{\prime
})(1|+\int_\Gamma dz^{\prime }V_{z^{\prime }}|1z^{\prime })(z^{\prime
}|-\int_{\overline{\Gamma }}dzV_z|z1)(z|+  \nonumber \\
&&-\int_\Gamma dz^{\prime }V_{z^{\prime }}|1)(1z^{\prime }|+\int_{\overline{%
\Gamma }}dz\int_\Gamma dz^{\prime }V_z|zz^{\prime })(1z^{\prime }|+\int_{%
\overline{\Gamma }}dzV_z|1)(z1|-  \nonumber \\
&&-\int_{\overline{\Gamma }}dz\int_\Gamma dz^{\prime }V_{z^{\prime
}}|zz^{\prime })(z1|+\int_{\overline{\Gamma }}dz\int_\Gamma dz^{\prime
}V_z|1z^{\prime })(zz^{\prime }|-\int_{\overline{\Gamma }}dz\int_\Gamma
dz^{\prime }V_{z^{\prime }}|z1)(zz^{\prime }|  \label{la17}
\end{eqnarray}

The superoperators $\stackrel{0}{\Bbb L}_{ext}$ and $\stackrel{1}{\Bbb L}%
_{ext}$ verify $(\rho |\stackrel{0}{\Bbb L}|O)=(\rho |\stackrel{0}{\Bbb L}%
_{ext}|O)$ and $(\rho |\stackrel{1}{\Bbb L}|O)=(\rho |\stackrel{1}{\Bbb L}%
_{ext}|O)$. The generalized eigenvalues of $\stackrel{0}{\Bbb L}_{ext}$ are
shown in Fig. 2. The form given in equations (\ref{la16}) and (\ref{la17})
for the superoperator ${\Bbb L}_{ext}=\stackrel{0}{\Bbb L}_{ext}+\stackrel{1}%
{\Bbb L}_{ext}$ can also be obtained directly from the definition ${\Bbb L}%
_{ext}O\doteq H_{\overline{\Gamma }}O-OH_\Gamma $, where $H_\Gamma $ was
defined in section II.

For the right eigenvalue problem ${\Bbb L}_{ext}|\Phi )=\lambda |\Phi )$ we
assume an expansion 
\[
|\Phi )=\,|\stackrel{0}{\Phi })+|\stackrel{1}{\Phi })+|\stackrel{2}{\Phi }%
)+\cdot \cdot \cdot \cdot \cdot ,\quad \lambda =\,\stackrel{0}{\lambda }+%
\stackrel{1}{\lambda }+\stackrel{2}{\lambda }+\cdot \cdot \cdot \cdot \cdot
, 
\]
with respect to the small interaction parameter that we suppose it is
contained in $\stackrel{1}{\Bbb L}$. The following set of equations is
obtained order by order 
\begin{eqnarray}
\left( \stackrel{0}{\lambda }-\stackrel{0}{\Bbb L}_{ext}\right) |\stackrel{0%
}{\Phi }{\Bbb )}\, &=&0  \label{lb1} \\
\left( \stackrel{0}{\lambda }-\stackrel{0}{\Bbb L}_{ext}\right) |\stackrel{1%
}{\Phi }{\Bbb )}\, &=&\left( \stackrel{1}{\Bbb L}_{ext}-\stackrel{1}{\lambda 
}\right) |\stackrel{0}{\Phi }{\Bbb )}  \label{lb2} \\
\left( \stackrel{0}{\lambda }-\stackrel{0}{\Bbb L}_{ext}\right) |\stackrel{2%
}{\Phi }{\Bbb )}\, &=&\left( \stackrel{1}{\Bbb L}_{ext}-\stackrel{1}{\lambda 
}\right) |\stackrel{1}{\Phi }{\Bbb )}-\stackrel{2}{\lambda }\,|\stackrel{0}{%
\Phi }{\Bbb )}  \label{lb3} \\
&&\cdot \cdot \cdot \cdot \cdot \cdot \cdot \cdot \cdot \cdot \cdot \cdot
\cdot \cdot  \nonumber \\
\left( \stackrel{0}{\lambda }-\stackrel{0}{\Bbb L}_{ext}\right) |\stackrel{n%
}{\Phi }{\Bbb )}\, &=&\left( \stackrel{1}{\Bbb L}_{ext}-\stackrel{1}{\lambda 
}\right) |\stackrel{n-1}{\Phi }{\Bbb )}-\stackrel{2}{\lambda }\,|\stackrel{%
n-2}{\Phi }{\Bbb )}-\cdot \cdot \cdot -\stackrel{n}{\lambda }\,|\stackrel{0}{%
\Phi }{\Bbb )}  \label{lb4} \\
&&\cdot \cdot \cdot \cdot \cdot \cdot \cdot \cdot \cdot \cdot \cdot \cdot
\cdot \cdot  \nonumber
\end{eqnarray}

For the left eigenvalue problem $(\Psi |{\Bbb L}_{ext}=\lambda (\Psi |$, the
expansion 
\[
(\Psi |=\,(\stackrel{0}{\Psi }|+(\stackrel{1}{\Psi }|+(\stackrel{2}{\Psi }%
|+\cdot \cdot \cdot \cdot \cdot ,\quad \lambda =\,\stackrel{0}{\lambda }+%
\stackrel{1}{\lambda }+\stackrel{2}{\lambda }+\cdot \cdot \cdot \cdot \cdot
, 
\]
gives

\begin{eqnarray}
{\Bbb (}\stackrel{0}{\Psi }|\left( \stackrel{0}{\lambda }-\stackrel{0}{\Bbb L%
}_{ext}\right) &=&0  \label{lb5} \\
{\Bbb (}\stackrel{1}{\Psi }|\left( \stackrel{0}{\lambda }-\stackrel{0}{\Bbb L%
}_{ext}\right) &=&{\Bbb (}\stackrel{0}{\Psi }|\left( \stackrel{1}{\Bbb L}%
_{ext}-\stackrel{1}{\lambda }\right)  \label{lb6} \\
{\Bbb (}\stackrel{2}{\Psi }|\left( \stackrel{0}{\lambda }-\stackrel{0}{\Bbb L%
}_{ext}\right) &=&{\Bbb (}\stackrel{1}{\Psi }|\left( \stackrel{1}{\Bbb L}%
_{ext}-\stackrel{1}{\lambda }\right) -\,{\Bbb (}\stackrel{0}{\Psi }|\,%
\stackrel{2}{\lambda }  \label{lb7} \\
&&\cdot \cdot \cdot \cdot \cdot \cdot \cdot \cdot \cdot \cdot \cdot \cdot 
\nonumber \\
{\Bbb (}\stackrel{n}{\Psi }|\left( \stackrel{0}{\lambda }-\stackrel{0}{\Bbb L%
}_{ext}\right) &=&{\Bbb (}\stackrel{n-1}{\Psi }|\left( \stackrel{1}{\Bbb L}%
_{ext}-\stackrel{1}{\lambda }\right) -\,{\Bbb (}\stackrel{n-2}{\Psi }|\,%
\stackrel{2}{\lambda }-\cdot \cdot \cdot -{\Bbb (}\stackrel{0}{\Psi }|\,%
\stackrel{n}{\lambda }  \label{lb8} \\
&&\cdot \cdot \cdot \cdot \cdot \cdot \cdot \cdot \cdot \cdot \cdot \cdot 
\nonumber
\end{eqnarray}

It is convenient to define the following projectors 
\begin{equation}
{\Bbb P}_0\doteq |1)(1|+\int d\omega |\omega )(\omega |,\quad {\Bbb P}_{%
\overline{\Gamma }\Gamma }\doteq \int_{\overline{\Gamma }}dz\int_\Gamma
dz^{\prime }|zz^{\prime })(zz^{\prime }|,\quad {\Bbb P}_{\overline{\Gamma }%
}\doteq \int_{\overline{\Gamma }}dz|z1)(z1|,\quad {\Bbb P}_\Gamma \doteq
\int_\Gamma dz^{\prime }|1z^{\prime })(1z^{\prime }|.  \label{lb9}
\end{equation}
These projectors commute with $\stackrel{0}{\Bbb L}_{ext}$ and also satisfy 
\[
{\Bbb I}_{ext}={\Bbb P}_0+{\Bbb P}_{\overline{\Gamma }\Gamma }+{\Bbb P}_{%
\overline{\Gamma }}+{\Bbb P}_\Gamma ,\qquad {\Bbb P}_0\stackrel{0}{\Bbb L}%
_{ext}=\stackrel{0}{\Bbb L}_{ext}{\Bbb P}_0=0.
\]
Therefore ${\Bbb P}_0$ is the projector on the invariant space of the time
evolution generated by the superoperator $\stackrel{0}{\Bbb L}_{ext}$. As in
the pure state case we will compute the discrete and the continuous spectrum.

\subsection{The discrete spectrum.}

We first consider equations (\ref{lb1})-(\ref{lb4}) with 
\begin{equation}
\stackrel{0}{\lambda }=0,\qquad |\stackrel{0}{\Phi })={\Bbb P}_0|\stackrel{0%
}{\Phi })=|1)(1|\stackrel{0}{\Phi })+\int d\omega |\omega )(\omega |%
\stackrel{0}{\Phi }).  \label{lb10}
\end{equation}
Multiplying equation (\ref{lb2}) by ${\Bbb P}_0$ and by ${\Bbb Q}_0\doteq 
{\Bbb I}_{ext}-{\Bbb P}_0$, we obtain 
\begin{eqnarray}
\stackrel{1}{\lambda }|\stackrel{0}{\Phi }) &=&{\Bbb P}_0\stackrel{0}{\Bbb L}%
_{ext}{\Bbb P}_0|\stackrel{0}{\Phi }),  \label{lb11} \\
{\Bbb Q}_0|\stackrel{1}{\Phi }) &=&\frac{(-1)}{{\Bbb Q}_0\stackrel{0}{\Bbb L}%
_{ext}{\Bbb Q}_0}{\Bbb Q}_0\stackrel{1}{\Bbb L}_{ext}|\stackrel{0}{\Phi }).
\label{lb12}
\end{eqnarray}
Equation (\ref{lb3}) gives 
\begin{eqnarray}
\stackrel{2}{\lambda }|\stackrel{0}{\Phi }) &=&\left[ {\Bbb P}_0\stackrel{1}%
{\Bbb L}_{ext}\frac{(-1)}{{\Bbb Q}_0\stackrel{0}{\Bbb L}_{ext}{\Bbb Q}_0}%
{\Bbb Q}_0\stackrel{1}{\Bbb L}_{ext}{\Bbb P}_0\right] |\stackrel{0}{\Phi }),
\label{lb13} \\
{\Bbb Q}_0|\stackrel{2}{\Phi }) &=&\frac{(-1)}{{\Bbb Q}_0\stackrel{0}{\Bbb L}%
_{ext}{\Bbb Q}_0}{\Bbb Q}_0\left[ \stackrel{1}{\Bbb L}_{ext}-\stackrel{1}{%
\lambda }\right] |\stackrel{1}{\Phi }).  \label{lb14}
\end{eqnarray}
From the definitions (\ref{lb9}) and (\ref{la17}) we have ${\Bbb P}_0%
\stackrel{0}{\Bbb L}_{ext}{\Bbb P}_0=0$ and therefore $\stackrel{1}{\lambda }%
=0$: the degeneration of the space expanded by the projector ${\Bbb P}_0$ is
not eliminated by first order corrections. The previous equations give no
information about ${\Bbb P}_0|\stackrel{n}{\Phi })$ ($n=1,2,...$), and we
make the usual choice $|\stackrel{n}{\Phi })={\Bbb Q}_0|\stackrel{n}{\Phi })$
($n=1,2,...$).

But the degeneration is partially removed with the second order correction.
Equation (\ref{lb13}), an eigenvalue problem for $\stackrel{2}{\lambda }$,
gives 
\begin{equation}
2\pi iV_\Omega ^2\,|1)(1|\stackrel{0}{\Phi })-2\pi iV_\Omega ^2\,|1)(\omega
=\Omega |\stackrel{0}{\Phi })=\stackrel{2}{\lambda }\,|1)(1|\stackrel{0}{%
\Phi })+\stackrel{2}{\lambda }\int d\omega |\omega )(\omega |\stackrel{0}{%
\Phi }).  \label{lb15}
\end{equation}
This equation has the solutions 
\begin{equation}
\stackrel{2}{\lambda }_d=2\pi iV_\Omega ^2,\quad |\stackrel{0}{\Phi }%
_d)=|1),\quad \quad \stackrel{2}{\lambda }_\omega =0,\quad |\stackrel{0}{%
\Phi }_\omega )=\delta (\omega -\Omega )|1)+|\omega ).  \label{lb16}
\end{equation}
We give in appendix A the details of these calculations.

For the left eigenvalue problem with 
\begin{equation}
\stackrel{0}{\lambda }=0,\qquad (\stackrel{0}{\Psi }|=(\stackrel{0}{\Psi }|%
{\Bbb P}_0=(\stackrel{0}{\Psi }|1)(1|+\int d\omega (\stackrel{0}{\Psi }%
|\omega )(\omega |,  \label{lb17}
\end{equation}
equations (\ref{lb5})-(\ref{lb8}) give 
\begin{eqnarray}
(\stackrel{0}{\Psi }|\stackrel{1}{\lambda } &=&(\stackrel{0}{\Psi }|({\Bbb P}%
_0\stackrel{0}{\Bbb L}_{ext}{\Bbb P}_0),  \label{lb18} \\
(\stackrel{1}{\Psi }|{\Bbb Q}_0 &=&(\stackrel{0}{\Psi }|\stackrel{1}{\Bbb L}%
_{ext}{\Bbb Q}_0\frac{(-1)}{{\Bbb Q}_0\stackrel{0}{\Bbb L}_{ext}{\Bbb Q}_0},
\label{lb19} \\
(\stackrel{0}{\Psi }|\stackrel{2}{\lambda } &=&(\stackrel{0}{\Psi }|\left[ 
{\Bbb P}_0\stackrel{1}{\Bbb L}_{ext}\frac{(-1)}{{\Bbb Q}_0\stackrel{0}{\Bbb L%
}_{ext}{\Bbb Q}_0}{\Bbb Q}_0\stackrel{1}{\Bbb L}_{ext}{\Bbb P}_0\right] ,
\label{lb20} \\
(\stackrel{2}{\Psi }|{\Bbb Q}_0 &=&(\stackrel{1}{\Psi }|\left[ \stackrel{1}%
{\Bbb L}_{ext}-\stackrel{1}{\lambda }\right] {\Bbb Q}_0\frac{(-1)}{{\Bbb Q}_0%
\stackrel{0}{\Bbb L}_{ext}{\Bbb Q}_0},  \label{lb21} \\
&&\cdot \cdot \cdot \cdot \cdot \cdot \cdot \cdot \cdot \cdot \cdot \cdot
\cdot \cdot \cdot \cdot \cdot \cdot \cdot \cdot \cdot  \nonumber
\end{eqnarray}

Once again, the degeneration is not removed by the first order corrections,
and $\stackrel{1}{\lambda }=0$ as a consequence of ${\Bbb P}_0\stackrel{0}%
{\Bbb L}_{ext}{\Bbb P}_0=0$. As there is no information about $(\stackrel{n}{%
\Psi }|{\Bbb P}_0$ for $n\geq 1$, the usual condition $(\stackrel{n}{\Psi }%
|=(\stackrel{n}{\Psi }|{\Bbb Q}_0$ can be imposed. For the eigenvalue
problem (\ref{lb20}) we obtain 
\begin{equation}
2\pi iV_\Omega ^2\,(\stackrel{0}{\Psi }|1)(1|-2\pi iV_\Omega ^2\,(\stackrel{0%
}{\Psi }|1)(\omega =\Omega |=\stackrel{2}{\lambda }(\stackrel{0}{\Psi }%
|1)(1|+\stackrel{2}{\lambda }\int d\omega (\stackrel{0}{\Psi }|\omega
)(\omega |.  \label{lb22}
\end{equation}
This equation has the solutions 
\begin{equation}
\stackrel{2}{\lambda }_d=2\pi iV_\Omega ^2,\quad (\stackrel{0}{\Psi }%
_d|=(1|-(\omega =\Omega |,\quad \quad \stackrel{2}{\lambda }_\omega =0,\quad
(\stackrel{0}{\Psi }_\omega |=(\omega |.  \label{lb23}
\end{equation}
(See appendix A for the details).

The right and left eigenvectors given in equations (\ref{lb16}) and (\ref
{lb23}) form a complete biorthogonal system for the ${\Bbb P}_0$ subspace,
i.e. 
\[
(\stackrel{0}{\Psi }_d|\stackrel{0}{\Phi }_d)=1,\qquad (\stackrel{0}{\Psi }%
_\omega |\stackrel{0}{\Phi }_{\omega ^{\prime }})=\delta (\omega -\omega
^{\prime }),\qquad (\stackrel{0}{\Psi }_d|\stackrel{0}{\Phi }_{\omega
^{\prime }})=(\stackrel{0}{\Psi }_\omega |\stackrel{0}{\Phi }_d)=0, 
\]
\[
{\Bbb P}_0=|\stackrel{0}{\Phi }_d)(\stackrel{0}{\Psi }_d|+\int d\omega |%
\stackrel{0}{\Phi }_\omega )(\stackrel{0}{\Psi }_\omega |. 
\]

The degeneration of the zero eigenvalue of $\stackrel{0}{\Bbb L}_{ext}$ have
been partially removed by the interaction: a proper eigenvalue $\lambda
_d\cong 2\pi iV_\Omega ^2$ with a single eigenvector is obtained. However,
the infinite degeneration of the eigenvalue $\lambda _\omega =0$ ($\omega
\in {\Bbb R}^{+}$) remains. This is a new feature that must be taken in
consideration: the infinitely degenerated eigenvalue corresponds to the
invariant states. As we see from equations (\ref{lb23}), these invariant
states are expanded, up to zero order, by the singular states $(\stackrel{0}{%
\Psi }_\omega |=(\omega |$. These invariant states only appear if we
introduce the singular structure for states and observables. The singular
structure is also necessary for the description of the decay of the unstable
discrete state. As we will see below, the imaginary eigenvalue $\lambda _d$
is associated with the exponential approximation of the decay process.

\subsection{The continuous spectrum.}

Let us consider the following generalized eigenvalues and eigenvectors of $%
\stackrel{0}{\Bbb L}_{ext}$ 
\begin{equation}
\stackrel{0}{\lambda }_{u1}=u-\Omega ,\qquad |\stackrel{0}{\Phi }%
_{u1})=|u1),\qquad (\stackrel{0}{\Psi }_{u1}|=(u1|,\qquad u\in \Gamma ,
\label{lc1}
\end{equation}
Equation (\ref{lb2}) gives 
\begin{equation}
(u-\Omega -\stackrel{0}{\Bbb L}_{ext})|\stackrel{1}{\Phi })=(\stackrel{1}%
{\Bbb L}_{ext}-\stackrel{1}{\lambda })|u1).  \label{lc2}
\end{equation}
When this equation is multiplied from the left by $(v1|$ we obtain 
\[
(u-v)(v1|\stackrel{1}{\Phi })=-\stackrel{1}{\lambda }\,\delta _{\overline{%
\Gamma }}(u-v),
\]
where $\delta _{\overline{\Gamma }}(u-v)$ is the ''$\delta $-function'' on
the curve $\overline{\Gamma }$ ($\int_{\overline{\Gamma }}du\,\delta _{%
\overline{\Gamma }}(u-v)\,f(u)=f(v)$). From this last equation we obtain $%
\stackrel{1}{\lambda }_{u1}=0$. The first order correction for the right
eigenvector can be obtained from equation (\ref{lc2}) 
\begin{equation}
|\stackrel{1}{\Phi }_{u1})=\frac{V_u}{u-\Omega }|1)-\int_\Gamma du^{\prime }%
\frac{V_{u^{\prime }}}{u^{\prime }-\Omega }|uu^{\prime }).  \label{lc3}
\end{equation}
From equation (\ref{lb6}) we obtain the first order correction of the left
eigenvector with the same eigenvalue 
\begin{equation}
(\stackrel{1}{\Psi }_{u1}|=\frac{V_u}{u-\Omega }[(1|-(u|]-\int_\Gamma
du^{\prime }\frac{V_{u^{\prime }}}{u^{\prime }-\Omega }(uu^{\prime }|.
\label{lc4}
\end{equation}
The second order correction to the eigenvalue can be obtained multiplying
equation (\ref{lb3}) by $(v1|$ from the left 
\begin{eqnarray*}
(u-v)(v1|\stackrel{2}{\Phi }_{u1}) &=&(v1|\stackrel{1}{\Bbb L}_{ext}|%
\stackrel{1}{\Phi })-\stackrel{2}{\lambda }_{u1}\,\delta _{\overline{\Gamma }%
}(u-v)= \\
&=&\left[ \int_\Gamma du^{\prime }\frac{V_{u^{\prime }}^2}{u^{\prime
}-\Omega }-\stackrel{2}{\lambda }_{u1}\right] \delta _{\overline{\Gamma }%
}(u-v)+\frac{V_uV_v}{u-\Omega }.
\end{eqnarray*}
From this equation we obtain 
\begin{equation}
\stackrel{2}{\lambda }_{u1}=\int_\Gamma du^{\prime }\frac{V_{u^{\prime }}^2}{%
u^{\prime }-\Omega }=\int_0^\infty d\omega \frac{V_\omega ^2}{\omega
-i0-\Omega }=i\pi V_\Omega ^2+\int_0^\infty d\omega \,V_\omega ^2\,{\cal P}%
\left( \frac 1{\omega -\Omega }\right) .  \label{lc5}
\end{equation}
Therefore, the generalized eigenvalues of $\stackrel{0}{\Bbb L}_{ext}$,
represented by points $z=u-\Omega $ ($u\in \overline{\Gamma }$) of the
complex plane, are shifted to the upper half plane by the interaction. The
corresponding eigenvalues of ${\Bbb L}_{ext}$ are the points $z=u-\Omega +%
\stackrel{2}{\lambda }_{u1}$, where $u\in \overline{\Gamma }$ and $\stackrel{%
2}{\lambda }_{u1}$ is given by equation (\ref{lc5}) (see Fig. 2).

We can also consider the following generalized eigenvalues and eigenvectors
of $\stackrel{0}{\Bbb L}_{ext}$%
\begin{equation}
\stackrel{0}{\lambda }_{1u^{\prime }}=\Omega -u^{\prime },\qquad |\stackrel{0%
}{\Phi }_{1u^{\prime }})=|1u^{\prime }),\qquad (\stackrel{0}{\Psi }%
_{1u^{\prime }}|=(1u^{\prime }|,\qquad u^{\prime }\in \Gamma ,  \label{lc5'}
\end{equation}
The perturbation method gives 
\begin{eqnarray}
\stackrel{1}{\lambda }_{1u^{\prime }} &=&0,  \nonumber \\
\stackrel{2}{\lambda }_{1u^{\prime }} &=&\int_{\overline{\Gamma }}du\frac{%
V_u^2}{\Omega -u}=i\pi V_\Omega ^2-\int_0^\infty d\omega \,V_\omega ^2\,%
{\cal P}\left( \frac 1{\omega -\Omega }\right) ,  \nonumber \\
|\stackrel{1}{\Phi }_{1u^{\prime }}) &=&\frac{V_{u^{\prime }}}{u^{\prime
}-\Omega }|1)-\int_{\overline{\Gamma }}du\frac{V_u}{u-\Omega }|uu^{\prime }),
\nonumber \\
(\stackrel{1}{\Psi }_{1u^{\prime }}| &=&\frac{V_{u^{\prime }}}{u^{\prime
}-\Omega }[(1|-(u^{\prime }|]-\int_{\overline{\Gamma }}du\frac{V_u}{u-\Omega 
}(uu^{\prime }|.  \label{lc6}
\end{eqnarray}
This result shows that the generalized eigenvalues of equation (\ref{lc5'})
are also shifted to the upper half plane by the interaction (see Fig. 2).

Finally, we can consider the following generalized eigenvalues and
eigenvectors of $\stackrel{0}{\Bbb L}_{ext}$ 
\begin{equation}
\stackrel{0}{\lambda }_{uu^{\prime }}=u-u^{\prime },\qquad |\stackrel{0}{%
\Phi }_{uu^{\prime }})=|uu^{\prime }),\qquad (\stackrel{0}{\Psi }%
_{uu^{\prime }}|=(uu^{\prime }|,\qquad u\in \overline{\Gamma },\;u^{\prime
}\in \Gamma ,  \label{lc6'}
\end{equation}
(see the shaded region in Fig.1). In this case the perturbed expansion gives 
\begin{eqnarray}
\stackrel{1}{\lambda }_{uu^{\prime }} &=&\stackrel{2}{\lambda }_{uu^{\prime
}}=0,  \nonumber \\
|\stackrel{1}{\Phi }_{uu^{\prime }}) &=&\frac{V_u}{u-\Omega }|1u^{\prime })+%
\frac{V_{u^{\prime }}}{u^{\prime }-\Omega }|u1),  \nonumber \\
(\stackrel{1}{\Psi }_{uu^{\prime }}| &=&\frac{V_u}{u-\Omega }(1u^{\prime }|+%
\frac{V_{u^{\prime }}}{u^{\prime }-\Omega }(u1|.  \label{lc6''}
\end{eqnarray}
There is no shift of this part of the generalized spectrum by the
interaction.

\subsection{Time evolution.}

In the previous subsection we obtained by a perturbation method a set of
generalized eigenvalues and eigenvectors of the superoperator ${\Bbb L}%
_{ext} $, which reduce to the complete biorthogonal set of $\stackrel{0}%
{\Bbb L}_{ext}$ when the parameter of the interaction goes to zero. The
generalized eigenvalues of ${\Bbb L}_{ext}$ are shown in Fig. 3. It seems
reasonable to assume that the obtained generalized eigenvalues and
eigenvectors of ${\Bbb L}_{ext}$, also form a complete biorthogonal set, at
least for small values of the interaction parameter \footnote{%
This fact can be explicitly verified up to second order for the generalized
eigenvectors obtained in sections IIIA and IIIB.}. If this is the case, the
time evolution operator is 
\begin{eqnarray}
\exp (i{\Bbb L}_{ext}t) &=&\int d\omega \,\exp (i\lambda _\omega t)\,|\Phi
_\omega )(\Psi _\omega |+\exp (i\lambda _dt)|\Phi _d)(\Psi _d|+\int_{%
\overline{\Gamma }}du\,\exp (i\lambda _{u1}t)|\Phi _{u1})(\Psi _{u1}|+ 
\nonumber \\
&&+\int_\Gamma du^{\prime }\,\exp (i\lambda _{1u^{\prime }}t)|\Phi
_{1u^{\prime }})(\Psi _{1u^{\prime }}|+\int_{\overline{\Gamma }%
}du\int_\Gamma du^{\prime }\,\exp (i\lambda _{uu^{\prime }}t)|\Phi
_{uu^{\prime }})(\Psi _{uu^{\prime }}|.  \label{ls1}
\end{eqnarray}
Up to second order, the eigenvalues are 
\begin{eqnarray}
\lambda _\omega &=&0,  \nonumber \\
\lambda _d &=&2\pi iV_\Omega ^2,  \nonumber \\
\lambda _{u1} &=&u-\Omega +i\pi V_\Omega ^2+\int_0^\infty d\omega \,V_\omega
^2\,{\cal P}\left( \frac 1{\omega -\Omega }\right) ,  \nonumber \\
\lambda _{1u^{\prime }} &=&\Omega -u^{\prime }+i\pi V_\Omega
^2-\int_0^\infty d\omega \,V_\omega ^2\,{\cal P}\left( \frac 1{\omega
-\Omega }\right) ,  \nonumber \\
\lambda _{uu^{\prime }} &=&u-u^{\prime }  \label{ls2}
\end{eqnarray}
and the approximated expressions for the eigenvectors are given in the
previous subsection.

The time evolution of the mean value of an observable $O$, given by 
\[
\langle O\rangle _{\rho _t}=(\rho _t|O)=(\rho _0|\exp (i{\Bbb L}t)|O), 
\]
can be easily computed using (\ref{ls1}). Let us consider the time evolution
of the unstable state given by $(\rho _0|=(1|$. If we call by $\varepsilon $
the small multiplicative interaction parameter included in $V_\omega $, we
can use the explicit approximated expressions of the eigenvectors to prove
that 
\begin{eqnarray*}
(1|\Phi _\omega )(\Psi _\omega |1) &=&O(\varepsilon ^2), \\
(1|\Phi _d)(\Psi _d|1) &=&1+O(\varepsilon ^2), \\
(1|\Phi _{u1})(\Psi _{u1}|1) &=&O(\varepsilon ^2), \\
(1|\Phi _{1u^{\prime }})(\Psi _{1u^{\prime }}|1) &=&O(\varepsilon ^2), \\
(1|\Phi _{uu^{\prime }})(\Psi _{uu^{\prime }}|1) &=&O(\varepsilon ^2),
\end{eqnarray*}
and therefore the survival probability of the unstable state which is
represented by the state functional $(\rho _0|=(1|$ (and also by the vector
state $|1\rangle $) is approximately given by 
\begin{equation}
(\rho _t|1)=(1|\exp (i{\Bbb L}t)|1)\cong \exp (-2\pi V_\Omega ^2t).
\label{ls3}
\end{equation}
This expression for the survival probability is valid if $\varepsilon ^2\ll
1 $ and $t\leq \varepsilon ^{-2}$.

Taking into account that 
\begin{eqnarray*}
(1|\Phi _{\omega ^{\prime }})(\Psi _{\omega ^{\prime }}|\omega ) &=&\delta
(\omega ^{\prime }-\Omega )\delta (\omega ^{\prime }-\omega )+O(\varepsilon
^2), \\
(1|\Phi _d)(\Psi _d|\omega ) &=&-\delta (\omega -\Omega )+O(\varepsilon ^2),
\\
(1|\Phi _{u1})(\Psi _{u1}|\omega ) &=&O(\varepsilon ^2), \\
(1|\Phi _{1u^{\prime }})(\Psi _{1u^{\prime }}|\omega ) &=&O(\varepsilon ^2),
\\
(1|\Phi _{uu^{\prime }})(\Psi _{uu^{\prime }}|\omega ) &=&O(\varepsilon ^2),
\end{eqnarray*}
for the initial unstable state $(\rho _0|=(1|$ we obtain 
\begin{equation}
(\rho _t|\omega )=(1|\exp (i{\Bbb L}t)|\omega )\cong \delta (\omega -\Omega
)\left[ 1-\exp (-2\pi V_\Omega ^2t)\right] ,  \label{ls4}
\end{equation}
also valid if $\varepsilon ^2\ll 1$ and $t\leq \varepsilon ^{-2}$. Equations
(\ref{ls3})-(\ref{ls4}) describes the transition of the unstable state $(1|$%
, into $(\omega =\Omega |$.

Therefore, the obtained set of generalized eigenvectors is a useful tool for
a complete description of the decay process. The pure state $|1\rangle $ is
given in this formalism by the state functional $(1|$, and the standard
approximated expression its the survival probability is reobtained in
equation (\ref{ls3}). But in addition we obtain in equation (\ref{ls4}) an
explicit expression for the by-products of the decay process.

\section{Conclusions.}

When faced with the problem of the spectral decomposition of the Hamiltonian 
\begin{eqnarray}
H &=&\stackrel{0}{H}+\stackrel{1}{H},  \nonumber \\
\stackrel{0}{H} &=&\Omega |1\rangle \langle 1|+\int_0^\infty d\omega
\,\omega |\omega \rangle \langle \omega |,\qquad \Omega \in {\Bbb R}^{+} 
\nonumber \\
\stackrel{1}{H} &=&\int_0^\infty d\omega \,V_\omega |\omega \rangle \langle
1|+\int_0^\infty d\omega \,\overline{V_\omega }|1\rangle \langle \omega
|+\int_0^\infty d\omega \int_0^\infty d\omega ^{\prime }V_{\omega \omega
^{\prime }}|\omega \rangle \langle \omega ^{\prime }|,  \label{c1}
\end{eqnarray}
we find that the superposition of the continuous and the discrete part of
the spectrum of $\stackrel{0}{H}$ makes impossible to apply the standard
methods of perturbation theory.

If we restrict the class of states (observables) of the system to the set
represented by vectors $\Phi $ ($\Psi $) for which $\langle \omega |\Phi
\rangle $ ($\langle \omega |\Psi \rangle $) has a well defined analytic
extension $\langle z|\Phi \rangle $ ($\langle z|\Psi \rangle $) to the lower
(upper) complex half plane, it is possible to decompose the amplitude $%
\langle \Psi |\Phi \rangle $ of the state $|\Phi \rangle $ to be observed in
the state $|\Psi \rangle $ in the form 
\begin{equation}
\langle \Psi |\Phi \rangle =\langle \Psi |I_\Gamma |\Phi \rangle ,\qquad
I_\Gamma \doteq |1\rangle \langle 1|+\int_\Gamma dz\,|z\rangle \langle z|,
\label{c2}
\end{equation}
being $\Gamma $ a curve in the lower complex half plane as shown in fig.1.

For the amplitude of the time evolved state $|\Phi _t\rangle =\exp
(-iHt)\,|\Phi \rangle $ to be observed in $|\Psi \rangle $, we also have 
\begin{equation}
\langle \Psi |\Phi _t\rangle =\langle \Psi |\exp (-iHt)\,|\Phi \rangle
=\langle \Psi |\exp (-iH_\Gamma t)\,|\Phi \rangle ,  \label{c3}
\end{equation}
where 
\begin{eqnarray}
H_\Gamma &=&\stackrel{0}{H}_\Gamma +\stackrel{1}{H}_\Gamma ,  \nonumber \\
\stackrel{0}{H}_\Gamma &\doteq &\Omega |1\rangle \langle 1|+\int_\Gamma
dz\,z|z\rangle \langle z|,  \nonumber \\
\stackrel{1}{H}_\Gamma &\doteq &\int_\Gamma dz\,V_z|z\rangle \langle
1|+\int_\Gamma dz\,\overline{V_{\overline{z}}}|1\rangle \langle
z|+\int_\Gamma dz\int_\Gamma dz^{\prime }V_{zz^{\prime }}|z\rangle \langle
z^{\prime }|.  \label{c4}
\end{eqnarray}

Starting from the right (left) eigenvectors $|1\rangle $ and $|z\rangle $ ($%
\langle 1|$ and $\langle z|$), which form a complete biorthogonal set for
the spectral decomposition of $\stackrel{0}{H}_\Gamma $, it is possible to
obtain a well defined perturbative algorithm to compute the corresponding
right (left) eigenvectors $|f_\Omega \rangle $ and $|f_z\rangle $ ($\langle 
\widetilde{f}_\Omega |$ and $\langle \widetilde{f}_z|$), a complete
biorthogonal set for the spectral decomposition of $H_\Gamma $, i.e. 
\begin{eqnarray}
I_\Gamma  &=&|f_\Omega \rangle \langle \widetilde{f}_\Omega |+\int_\Gamma
du\,|f_u\rangle \langle \widetilde{f}_u|  \label{c5} \\
H_\Gamma  &=&\lambda _\Omega \,|f_\Omega \rangle \langle \widetilde{f}%
_\Omega |+\int_\Gamma du\,\lambda _u\,|f_u\rangle \langle \widetilde{f}_u|,
\label{c6}
\end{eqnarray}
\begin{equation}
\langle \widetilde{f}_\Omega |f_\Omega \rangle =1,\quad \langle \widetilde{f}%
_u|f_{u^{\prime }}\rangle =\delta _\Gamma (u-u^{\prime }),\quad \langle 
\widetilde{f}_\Omega |f_u\rangle =\langle \widetilde{f}_u|f_\Omega \rangle
=0.  \label{c7}
\end{equation}

For the Hamiltonian $H_\Gamma $ of the general form given in equations (\ref
{c4}), the orthogonality and completeness relations (equations (\ref{c7})
and (\ref{c5})) can be verified by explicit calculations up to any finite
order in the perturbation. For the Friedrichs model (a special case of the
Hamiltonian given in equation (\ref{c1})), the solutions of the eigenvalue
problem obtained by the perturbation alghorithm coincide with the exact
solutions obtained by Sudarshan C.B.Chiu, V.Gorini \cite{3}. The same
complex spectral decomposition for the Friedrichs model was also obtained by
T.Petrosky, I.Prigogine and S.Tasaki \cite{3b}, with a perturbative method
using a ''time ordering rule'' by which an small imaginary part is included
to avoid small denominators, with a different sign according to the type of
transition involved. In the perturbative algorithm presented in this paper,
there is no need of additional rules to avoid the singularities due to
resonances between discrete and continuous parts of the unperturbed
spectrum. When states and observables are restricted to have suitable
analytic properties, the deformation of the integrals over the continuous
part of the spectrum to a curve in the complex lower half plane, eliminate
the continuous-discrete resonances. After the expressions for the
eigenvalues and eigenvectors are obtained up to de desired order, the
complex contour of integration can be deformed back to the real axis. In
this way, the ''time ordering rule'' of reference \cite{3b} can be deduced
from the analytic extension properties of states and observables.

If the transition probability 
\begin{equation}
P_t=\left| \langle \Psi |\exp (-iHt)|\Phi \rangle \right| ^2=\left| \exp
(-i\,\lambda _\Omega \,t)\langle \Psi |f_\Omega \rangle \langle \widetilde{f}%
_\Omega |\Phi \rangle +\int_\Gamma dz\,\exp (-i\,z\,t)\langle \Psi
|f_z\rangle \langle \widetilde{f}_z|\Phi \rangle \right| ^2,  \label{c8}
\end{equation}
is obtained through eigenvalues and eigenvectors computed up to n-th order,
the necessary conditions for a good approximation of $P_t$ are $\varepsilon
^{n+1}\ll 1$ and $\varepsilon ^{n+1}t\ll 1$ where $\varepsilon $ is the
small interaction parameter. Then, even for small interactions, it is only
for not too large times that we can obtain a good approximation of $P_t$.

An important advantage of the spectral decomposition in terms of generalized
eigenvectors with complex eigenvalues is that it can be obtained with a well
defined perturbation algorithm. The fact that this spectral decomposition is
only feasible for states $|\Phi \rangle $ and observables $|\Psi \rangle $
with the analytic properties mentioned above is not a limitation. When we
have to model a state or an observable from the finite amount of information
coming from experimental data it is always possible to choose $\langle
\omega |\Phi \rangle $ ($\langle \omega |\Psi \rangle $) with well defined
analytic extensions to the lower (upper) complex half plane. Moreover, we
are usually free to choose upper analytic extensions for the states and
lower analytic extensions for the observables \cite{2}. For this choice, we
have 
\begin{equation}
P_t=\left| \langle \Psi |\exp (-iHt)|\Phi \rangle \right| ^2=\left| \exp
(-i\,\overline{\lambda }_\Omega \,t)\langle \Phi |\widetilde{f}_\Omega
\rangle \langle f_\Omega |\Psi \rangle +\int_{\overline{\Gamma }}dz\,\exp
(-i\,z\,t)\,\langle \Phi |\widetilde{f}_z\rangle \langle f_z|\Psi \rangle
\right| ^2,  \label{c9}
\end{equation}
where $|\widetilde{f}_\Omega \rangle $ and $|\widetilde{f}_z\rangle $ ($%
\langle f_\Omega |$ and $\langle f_z|$) are right (left) generalized
eigenvectors of $H_{\overline{\Gamma }}$, being $\overline{\Gamma }$ a curve
in the upper half of the complex plane.

For states and observables having {\it both} analytic extensions to the
upper and lower half plane near the real axis, both expressions (\ref{c8})
and (\ref{c9}) can be used to obtain $P_t$. However, equation (\ref{c9})
would be a bad choice for $t>0$, as the positive imaginary part of the
eigenvalues will give a {\it well defined} $P_t$ in terms of unbounded
factors (the time evolution would include exponentially growing terms).

We also considered in this paper the class of observables represented by
operators of the form 
\begin{equation}
|O)=O_1|1)+\int_0^\infty d\omega \,O_\omega \,|\omega )+\int_0^\infty
d\omega \int_0^\infty d\omega ^{\prime }\,O_{\omega \omega ^{\prime
}}\,|\omega \omega ^{\prime })+\int_0^\infty d\omega \,O_{\omega 1}\,|\omega
1)+\int_0^\infty d\omega ^{\prime }\,O_{1\omega ^{\prime }}\,|1\omega
^{\prime }),  \label{c11}
\end{equation}
where the second term of the right hand side is a singular operator, and 
\begin{equation}
|1)\doteq |1\rangle \langle 1|,\qquad |\omega )\doteq |\omega \rangle
\langle \omega |,\qquad |\omega \omega ^{\prime })\doteq |\omega \rangle
\langle \omega ^{\prime }|,\qquad |\omega 1)\doteq |\omega \rangle \langle
1|,\qquad |1\omega ^{\prime })\doteq |1\rangle \langle \omega ^{\prime }|.
\label{c12}
\end{equation}

Following the work of references \cite{4}, \cite{5} and \cite{6}, we
considered the states as functionals $\rho $ acting on observables $O$ to
give the mean value of the observable in the state ($\langle O\rangle _\rho
\doteq (\rho |O)$). The states can be expressed as linear combinations of
the functionals defined by 
\begin{equation}
(1|O)=O_1,\quad (\omega |O)=O_\omega ,\quad (\omega \omega ^{\prime
}|O)=O_{\omega \omega ^{\prime }},\quad (\omega 1|O)=O_{\omega 1},\quad
(1\omega ^{\prime }|O)=O_{1\omega ^{\prime }}.  \label{c13}
\end{equation}

The Liouville-Von Newmann superoperators $\stackrel{0}{\Bbb L}=[\stackrel{0}{%
H},\;]$ and $\stackrel{1}{\Bbb L}=[\stackrel{1}{H},\;]$ can be expanded in
terms of the generalized observables and states given in equations (\ref{c12}%
) and (\ref{c13}). For the Friedrichs model we obtain 
\begin{eqnarray}
\stackrel{0}{\Bbb L} &=&\int d\omega ^{\prime }(\Omega -\omega ^{\prime
})|1\omega ^{\prime })(1\omega ^{\prime }|+\int d\omega (\omega -\Omega
)|\omega 1)(\omega 1|+\int d\omega \int d\omega ^{\prime }(\omega -\omega
^{\prime })|\omega \omega ^{\prime })(\omega \omega ^{\prime }|,  \label{c14}
\\
\stackrel{1}{\Bbb L} &=&\int d\omega V_\omega \left[ |\omega 1)-|1\omega
)\right] (1|+\int d\omega V_\omega \left[ |1\omega )-|\omega 1)\right]
(\omega |+\int d\omega \left[ -V_\omega |1)+\int d\omega ^{\prime }V_{\omega
^{\prime }}|\omega ^{\prime }\omega )\right] (1\omega |+  \nonumber \\
&&+\int d\omega \left[ V_\omega |1)-\int d\omega ^{\prime }V_{\omega
^{\prime }}|\omega \omega ^{\prime })\right] (\omega 1|+\int d\omega \int
d\omega ^{\prime }\left[ V_\omega |1\omega ^{\prime })-V_{\omega ^{\prime
}}|\omega 1)\right] (\omega \omega ^{\prime }|.  \label{c15}
\end{eqnarray}

Assuming suitable analytic properties for $\rho $ and $O$, the eigenvalue
problem for ${\Bbb L}=\stackrel{0}{\Bbb L}+\stackrel{1}{\Bbb L}$ can be
transformed into the eigenvalue problem for ${\Bbb L}_{ext}$, obtained from $%
{\Bbb L}$ replacing the integrals $\int_0^\infty d\omega $ and $%
\int_0^\infty d\omega ^{\prime }$ by $\int_{\overline{\Gamma }}dz$ and $%
\int_\Gamma dz^{\prime }$, being $\overline{\Gamma }$ ($\Gamma $) a curve in
the upper (lower) complex half plane. Starting from the right and left
generalized eigenvectors of $\stackrel{0}{\Bbb L}_{ext}$, a well defined
perturbation algorithm can be used to obtain a complete biorthogonal set of
generalized eigenvectors of ${\Bbb L}_{ext}$%
\[
{\Bbb L}_{ext}=\int d\omega \,\lambda _\omega \,|\Phi _\omega )(\Psi _\omega
|+\lambda _d|\Phi _d)(\Psi _d|+\int_{\overline{\Gamma }}du\,\lambda
_{u1}|\Phi _{u1})(\Psi _{u1}|+\int_\Gamma du^{\prime }\,\lambda _{1u^{\prime
}}|\Phi _{1u^{\prime }})(\Psi _{1u^{\prime }}|+\int_{\overline{\Gamma }%
}du\int_\Gamma du^{\prime }\,\lambda _{uu^{\prime }}|\Phi _{uu^{\prime
}})(\Psi _{uu^{\prime }}|. 
\]
\[
{\Bbb I}_{ext}=\int d\omega \,|\Phi _\omega )(\Psi _\omega |+|\Phi _d)(\Psi
_d|+\int_{\overline{\Gamma }}du\,|\Phi _{u1})(\Psi _{u1}|+\int_\Gamma
du^{\prime }\,|\Phi _{1u^{\prime }})(\Psi _{1u^{\prime }}|+\int_{\overline{%
\Gamma }}du\int_\Gamma du^{\prime }\,|\Phi _{uu^{\prime }})(\Psi
_{uu^{\prime }}|. 
\]

The generalized eigenvalues of $\stackrel{0}{\Bbb L}_{ext}$ and ${\Bbb L}%
_{ext}$ are shown in Figs. 2 and 3. The time evolution of the mean value for
any observable $O$ of the form given in equation (\ref{c11}) can be obtained
in terms of these generalized eigenvalues and eigenvectors. For example, it
is possible to obtain the time evolution of the unstable state $\rho _0=(1|$%
. If $\varepsilon ^2\ll 1$ and $t\leq \varepsilon ^{-2}$ (being $\varepsilon 
$ the interaction parameter), we obtain 
\[
(\rho _t|=(1|\exp (i{\Bbb L}t)\cong \exp (-2\pi V_\Omega ^2t)(1|+\left[
1-\exp (-2\pi V_\Omega ^2t)\right] (\omega =\Omega |. 
\]

Therefore, the obtained set of generalized eigenvectors is a useful tool for
a complete description of the decay process. The pure state $|1\rangle $ is
given in this formalism by the state functional $(1|$, and the standard
approximated expression for its survival probability is reobtained. But in
addition we obtain an explicit expression for the by-products of the decay
process.

The generalized spectral decomposition is useful to describe the time
evolution of the physical state functionals, because the contribution of the
resonances is incorporated through the exponential dumping factors. However,
it is important to understand that not all the generalized left eigenvectors
of the Liouville-Von Newmann superoperator ${\Bbb L}$ involved in the
spectral decomposition can have an independent physical meaning: Let us
consider $(\rho |=(\Psi _\lambda |$, where $(\Psi _\lambda |$ is a left
generalized eigenvector of ${\Bbb L}$ with eigenvalue $\lambda \neq 0$.
Therefore we have $(\Psi _\lambda |{\Bbb L}=\lambda (\Psi _\lambda |$.
Acting with both sides of this equation on the identity operator $I$ we
obtain 
\[
\lambda (\Psi _\lambda |I)=(\Psi _\lambda |{\Bbb L}I)=(\Psi _\lambda
|HI-IH)=(\Psi _\lambda |H-H)=0.
\]
If $\lambda \neq 0$, $(\Psi _\lambda |I)=0$ and the total probability
condition $(\rho |I)=1$ is not verified by $(\Psi _\lambda |$. Therefore, it
can not be a physical state.

\begin{center}
{\bf ACKNOWLEDGMENTS}
\end{center}

This work was partially supported by Grant No. CI1-CT94-0004 of the European
Community, Grant No. PID-0150 of CONICET (National Research Council of
Argentina), Grant No. EX-198 of Buenos Aires University, Grant No. 12217/1
of Fundaci\'{o}n Antorchas, and also a Grant from the Foundation pour la
Recherche Foundamentale OLAM.

\appendix

\section{Second order eigenvalues for the discrete spectrum.}

As we showed in section II, $\stackrel{0}{\Bbb L}_{ext}$ has a zero
eigenvalue with infinite degeneration. This degeneration remains unchanged
up to first order but it is partially removed in the second order
approximation.

For the right eigenvalue problem, the second order corrections $\stackrel{2}{%
\lambda }$ to the eigenvalues is given by equation (\ref{lb15}) 
\begin{equation}
2\pi iV_\Omega ^2\,|1)(1|\stackrel{0}{\Phi })-2\pi iV_\Omega ^2\,|1)(\omega
=\Omega |\stackrel{0}{\Phi })=\stackrel{2}{\lambda }\,|1)(1|\stackrel{0}{%
\Phi })+\stackrel{2}{\lambda }\int d\omega |\omega )(\omega |\stackrel{0}{%
\Phi }),  \label{lA1}
\end{equation}
which is equivalent to the set of equations 
\begin{eqnarray}
\stackrel{2}{\lambda }\,(1|\stackrel{0}{\Phi }) &=&2\pi iV_\Omega ^2\,(1|%
\stackrel{0}{\Phi })-2\pi iV_\Omega ^2\,(\omega =\Omega |\stackrel{0}{\Phi })
\label{lA2} \\
\stackrel{2}{\lambda }(\omega |\stackrel{0}{\Phi }) &=&0,\qquad \omega \in 
{\Bbb R}^{+}.  \label{lA3}
\end{eqnarray}
If $\stackrel{2}{\lambda }\neq 0$, equations (\ref{lA3}) (\ref{lA2}) give $%
(\omega |\stackrel{0}{\Phi })=0$ and $\stackrel{2}{\lambda }\,(1|\stackrel{0%
}{\Phi })=2\pi iV_\Omega ^2\,(1|\stackrel{0}{\Phi })$. Therefore 
\begin{equation}
\stackrel{2}{\lambda }_d=2\pi iV_\Omega ^2,\qquad |\stackrel{0}{\Phi }%
_d)=|1).  \label{lA4}
\end{equation}
If $\stackrel{2}{\lambda }=0$, equation (\ref{lA3}) give no condition on $%
(\omega |\stackrel{0}{\Phi })$, so we can choose $(\omega |\stackrel{0}{\Phi 
})=\delta (\omega -\widetilde{\omega })$. Replacing in (\ref{lA2}) we obtain 
$(1|\stackrel{0}{\Phi })=\delta (\Omega -\widetilde{\omega })$, and
therefore 
\begin{equation}
\stackrel{2}{\lambda }_{\widetilde{\omega }}=0,\qquad |\stackrel{0}{\Phi }_{%
\widetilde{\omega }})=|\widetilde{\omega })+\delta (\widetilde{\omega }%
-\Omega )|1).  \label{lA5}
\end{equation}

For the left eigenvectors, the second order correction to the eigenvalues is
given by equation (\ref{lb22}) 
\begin{equation}
2\pi iV_\Omega ^2\,(\stackrel{0}{\Psi }|1)(1|-2\pi iV_\Omega ^2\,(\stackrel{0%
}{\Psi }|1)(\omega =\Omega |=\stackrel{2}{\lambda }(\stackrel{0}{\Psi }%
|1)(1|+\stackrel{2}{\lambda }\int d\omega (\stackrel{0}{\Psi }|\omega
)(\omega |,  \label{lA6}
\end{equation}
or equivalently 
\begin{eqnarray}
\stackrel{2}{\lambda }(\stackrel{0}{\Psi }|1) &=&2\pi iV_\Omega ^2\,(%
\stackrel{0}{\Psi }|1),  \label{lA7} \\
\stackrel{2}{\lambda }(\stackrel{0}{\Psi }|\omega ) &=&-2\pi iV_\Omega ^2\,(%
\stackrel{0}{\Psi }|1)\,\delta (\omega -\Omega ).  \label{lA8}
\end{eqnarray}

If $\stackrel{2}{\lambda }\neq 0$, equation (\ref{lA7}) gives $\stackrel{2}{%
\lambda }_d=2\pi iV_\Omega ^2$ and no condition on $(\stackrel{0}{\Psi }|1)$%
. If we choose $(\stackrel{0}{\Psi }|1)=1$, equation (\ref{lA8}) implies $(%
\stackrel{0}{\Psi }_d|\omega )=-\delta (\omega -\Omega )$. Therefore 
\begin{equation}
\stackrel{2}{\lambda }_d=2\pi iV_\Omega ^2,\qquad (\stackrel{0}{\Psi }%
_d|=(1|-\int d\omega \delta (\omega -\Omega )(\omega |=(1|-(\omega =\Omega |
\label{lA9}
\end{equation}
If $\stackrel{2}{\lambda }=0$, equation (\ref{lA7}) implies $(\stackrel{0}{%
\Psi }|1)=0$, while equation (\ref{lA8}) gives no condition on $(\stackrel{0%
}{\Psi }|\omega )$. We can choose 
\begin{equation}
\stackrel{2}{\lambda }_\omega =0,\qquad (\stackrel{0}{\Psi }_\omega
|=(\omega |.  \label{lA10}
\end{equation}
We have obtained the eigenvalues and eigenvectors quoted in equations (\ref
{lb16}) and (\ref{lb23}). The normalization constants have been chosen to
satisfy the orthogonality conditions 
\begin{equation}
(\stackrel{0}{\Psi }_d|\stackrel{0}{\Phi }_d)=1,\qquad (\stackrel{0}{\Psi }%
_\omega |\stackrel{0}{\Phi }_{\omega ^{\prime }})=\delta (\omega -\omega
^{\prime }),\qquad (\stackrel{0}{\Psi }_d|\stackrel{0}{\Phi }_\omega )=(%
\stackrel{0}{\Psi }_\omega |\stackrel{0}{\Phi }_d)=0,  \label{lA11}
\end{equation}
as can be easily verified using equations (\ref{lA4}), (\ref{lA5}), (\ref
{lA9}) and (\ref{lA10}). It is also straighforward to verify that these
eigenvectors form a complete set to expand the subspace generated by ${\Bbb P%
}_0$, i.e. 
\begin{equation}
{\Bbb P}_0=|1)(1|+\int d\omega |\omega )(\omega |=|\stackrel{0}{\Phi }_d)(%
\stackrel{0}{\Psi }_d|+\int d\omega |\stackrel{0}{\Phi }_\omega )(\stackrel{0%
}{\Psi }_\omega |.  \label{lA12}
\end{equation}

\end{document}